\documentclass[10pt,nobalancelastpage,twocolumn,floatfix,superscriptaddress,aps,prb]{revtex4-2}

\usepackage{microtype}
\usepackage[hidelinks,colorlinks,citecolor=blue,linkcolor=black,urlcolor=blue]{hyperref}
\usepackage{graphicx}% Include figure files
\usepackage{amsmath,amssymb,amsfonts}
\usepackage[capitalize]{cleveref}
\usepackage{xspace}
\usepackage{dsfont}
\usepackage[caption=false]{subfig}
\usepackage{floatrow}
\usepackage{tabularx}
\usepackage{booktabs}
\usepackage{tikz}
\usepackage{mathtools}
\usepackage{colortbl}
\usepackage{placeins}
\usepackage{siunitx}

\graphicspath{{./}{../plots/}}

\let\originalleft\left
\let\originalright\right
\renewcommand{\left}{\mathopen{}\mathclose\bgroup\originalleft}
\renewcommand{\right}{\aftergroup\egroup\originalright}

\definecolor{my_orange}{HTML}{F5793A}
\DeclareRobustCommand\filledDot{\tikz{\fill[my_orange] (1ex,1ex) circle (1ex);}\xspace}
\DeclareRobustCommand\unfilledDot{\tikz{\draw[my_orange,line width=0.6pt] (1ex,1ex) circle (1ex);}\xspace}

\definecolor{LightRed}{HTML}{FFE2DC}

\newsavebox{\measurebox}

\begin{document}

\title{Measurement-induced criticality and entanglement clusters:\\ a study of 1D and 2D Clifford circuits}

\author{Oliver Lunt}
\email{oliver.lunt.17@ucl.ac.uk}
\affiliation{Department of Physics, University College London, Gower Street, London, WC1E 6BT}

\author{Marcin Szyniszewski}
\email{mszynisz@gmail.com}
\affiliation{Department of Physics, University College London, Gower Street, London, WC1E 6BT}

\author{Arijeet Pal}
\email{a.pal@ucl.ac.uk}
\affiliation{Department of Physics, University College London, Gower Street, London, WC1E 6BT}

\date{\today}

\begin{abstract}

Entanglement transitions in quantum dynamics present a novel class of phase transitions in non-equilibrium systems. When a many-body quantum system undergoes unitary evolution interspersed with monitored random measurements, the steady-state can exhibit a phase transition between volume- and area-law entanglement. 
There is a correspondence between measurement-induced transitions in non-unitary quantum circuits in $d$ spatial dimensions and classical statistical mechanical models in $d+1$ dimensions. In certain limits these models map to percolation, but there is analytical and numerical evidence to suggest that away from these limits the universality class should generically be distinct from percolation.
Intriguingly, despite these arguments, numerics on 1+1D qubit circuits give \textit{bulk} exponents which are nonetheless close to those of 2D percolation, with some possible differences in surface behavior.
In the first part of this work we explore the critical properties of 2+1D Clifford circuits. In the bulk, we find many properties suggested by the percolation picture, including several matching bulk exponents, and an inverse power-law for the critical entanglement growth, $S(t,L) \sim L(1 - a/t)$, which saturates to an area-law.
We then utilize a graph-state based algorithm to analyze in 1+1D and 2+1D the critical properties of entanglement clusters in the steady state. We show that in a model with a simple geometric map to percolation --- the projective transverse field Ising model --- these entanglement clusters are governed by percolation surface exponents. However, in the Clifford models we find large deviations in the cluster exponents from those of surface percolation, highlighting the breakdown of any possible geometric map to percolation.
Given the evidence for deviations from the percolation universality class, our results raise the question of why nonetheless many bulk properties behave similarly to those of percolation.

\end{abstract}

\maketitle
    
\section{Introduction}

Recent years have seen the exciting discovery of novel non-equilibrium phases of matter in many-body quantum systems. Quantum entanglement provides a natural framework for the taxonomy of these non-equilibrium phases. A prominent example of a non-equilibrium phase transition is the many-body localization (MBL) transition~\cite{baskoMetalInsulatorTransition2006, Gornyi2005, palManybodyLocalizationPhase2010,nandkishoreManyBodyLocalizationThermalization2015,abaninManybodyLocalizationThermalization2018,parameswaranManybodyLocalizationSymmetry2018}, in which the energy eigenstates switch from area-law entanglement in the MBL phase to volume-law in the chaotic phase. This singular change in the entanglement scaling means that the MBL transition is an example of an \textit{entanglement transition}.

\begin{figure}[b!]
    \centering
    \includegraphics[width=\columnwidth]{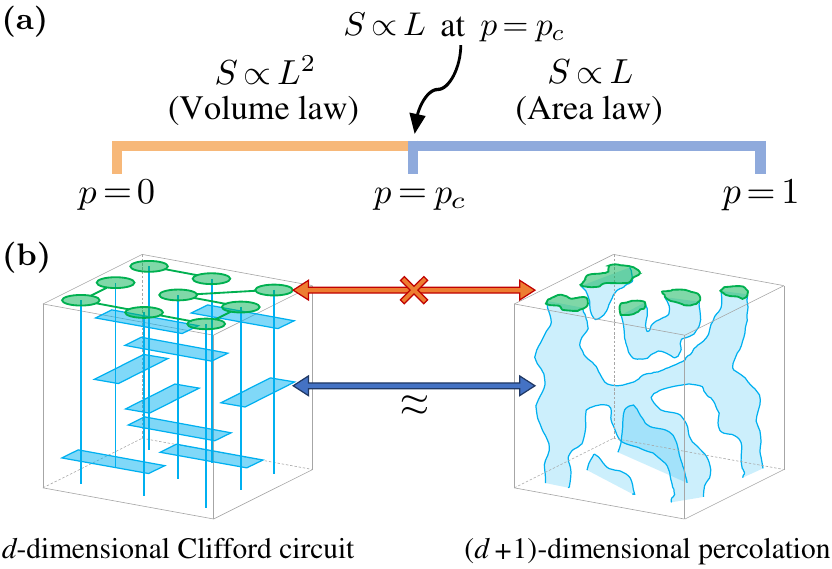}
    \caption{\textbf{(a)} Phase diagram for the measurement-induced transition in 2+1D local random Clifford circuits. For measurement probabilities $p<p_{c}$ the steady state exhibits volume-law entanglement, while for $p \geq p_{c}$ the steady state is area-law entangled. The entanglement transition and the purification transition coincide. \textbf{(b)} The critical point of a $d$-dimensional circuit appears to be described by bulk exponents from the $(d+1)$D percolation. However, entanglement cluster exponents do not match the percolation surface exponents.}
    \label{fig:summary}
\end{figure}

For systems without energy conservation, random unitary circuits have served as effective toy models of many-body quantum chaos~\cite{Nahum2017_RUD, Chan2018, Bertini2019,farshiTimeperiodicDynamicsGenerates2020}. With the advent of noisy intermediate scale quantum (NISQ) devices~\cite{preskillQuantumComputingNISQ2018}, the dynamics of pseudo-random unitary circuits can now be realized in experimental platforms, including superconducting qubits \cite{arute2019qSupreme} and trapped ions \cite{landsman2019scrambling}. 
A many-body quantum system undergoing chaotic unitary time evolution will typically thermalize, leading to volume-law entanglement in the steady state~\cite{dalessioQuantumChaosEigenstate2016}. However, this thermalization and the concomitant volume law can be destroyed if the time evolution becomes \textit{non-unitary} due to randomly interspersed measurements. The steady state conditioned on the measurement outcomes can then exhibit a phase transition between volume- and area-law entanglement as a function of the measurement rate, leading to the notion of \textit{measurement-induced} transitions~\cite{liQuantumZenoEffect2018,chanUnitaryprojectiveEntanglementDynamics2019,skinnerMeasurementInducedPhaseTransitions2019,szyniszewskiEntanglementTransitionVariablestrength2019,liMeasurementdrivenEntanglementTransition2019,nappEfficientClassicalSimulation2019,zabaloCriticalPropertiesMeasurementinduced2020,fanSelfOrganizedErrorCorrection2020,gullansDynamicalPurificationPhase2020,baoTheoryPhaseTransition2020,jianMeasurementinducedCriticalityRandom2020,liConformalInvarianceQuantum2020,shtankoClassicalModelsEntanglement2020,lavasaniMeasurementinducedTopologicalEntanglement2020,sangMeasurementProtectedQuantum2020,szyniszewskiUniversalityEntanglementTransitions2020,zhangNonuniversalEntanglementLevel2020,choiQuantumErrorCorrection2020,turkeshiMeasurementinducedCriticalityDimensional2020,gullansScalableProbesMeasurementInduced2020,nahumMeasurementEntanglementPhase2020,caoEntanglementFermionChain2019,tangMeasurementinducedPhaseTransition2020,gotoMeasurementInducedTransitionsEntanglement2020,albertonTrajectoryDependentEntanglement2020,luntMeasurementinducedEntanglementTransitions2020,langEntanglementTransitionProjective2020,chenEmergentConformalSymmetry2020,liuNonunitaryDynamicsSachdevYeKitaev2020,fujiMeasurementinducedQuantumCriticality2020,ippolitiEntanglementPhaseTransitions2020,vanregemortelEntanglementEntropyScaling2020,aharonovQuantumClassicalPhase2000,vijayMeasurementDrivenPhaseTransition2020,nahumEntanglementDynamicsDiffusionannihilation2020,liStatisticalMechanicsQuantum2020,rossiniMeasurementinducedDynamicsManybody2020,gullansScalableProbesMeasurementInduced2020,gullansQuantumCodingLowdepth2020,fidkowskiHowDynamicalQuantum2020,maimbourgBathinducedZenoLocalization2020,iaconisMeasurementinducedPhaseTransitions2020,ippolitiPostselectionfreeEntanglementDynamics2020,lavasaniTopologicalOrderCriticality2020,sangEntanglementNegativityMeasurementInduced2020,shiEntanglementNegativityCritical2020,gopalakrishnanEntanglementPurificationTransitions2020}.

These measurement-induced transitions occur in a wide variety of models, including random circuits~\cite{liQuantumZenoEffect2018,chanUnitaryprojectiveEntanglementDynamics2019,skinnerMeasurementInducedPhaseTransitions2019,szyniszewskiEntanglementTransitionVariablestrength2019,liMeasurementdrivenEntanglementTransition2019,nappEfficientClassicalSimulation2019,zabaloCriticalPropertiesMeasurementinduced2020,fanSelfOrganizedErrorCorrection2020,baoTheoryPhaseTransition2020,jianMeasurementinducedCriticalityRandom2020,liConformalInvarianceQuantum2020,shtankoClassicalModelsEntanglement2020,lavasaniMeasurementinducedTopologicalEntanglement2020,sangMeasurementProtectedQuantum2020,gullansDynamicalPurificationPhase2020,szyniszewskiUniversalityEntanglementTransitions2020,zhangNonuniversalEntanglementLevel2020,choiQuantumErrorCorrection2020,turkeshiMeasurementinducedCriticalityDimensional2020,gullansScalableProbesMeasurementInduced2020,nahumMeasurementEntanglementPhase2020}, Hamiltonian systems~\cite{caoEntanglementFermionChain2019,tangMeasurementinducedPhaseTransition2020,gotoMeasurementInducedTransitionsEntanglement2020,albertonTrajectoryDependentEntanglement2020,luntMeasurementinducedEntanglementTransitions2020,langEntanglementTransitionProjective2020,chenEmergentConformalSymmetry2020,liuNonunitaryDynamicsSachdevYeKitaev2020,fujiMeasurementinducedQuantumCriticality2020}, and measurement-only models~\cite{ippolitiEntanglementPhaseTransitions2020,vanregemortelEntanglementEntropyScaling2020,lavasaniMeasurementinducedTopologicalEntanglement2020,lavasaniTopologicalOrderCriticality2020}, and exhibit universal behavior. However, the determination of the relevant universality classes has proved to be a subtle issue. In certain 1+1D systems there is a `dimensional correspondence', where the measurement-induced transition in the 1+1D quantum system corresponds to an ordering transition in a 2+0D statistical mechanical model. Through these models, it has become clear that there is an important link between measurement-induced transitions and classical percolation, but the precise nature of this relationship is still unclear. For example, for 1+1D Haar-random circuits there are two distinct mappings to 2D percolation: one for the $(n=0)$-R\'{e}nyi entropy (Hartley entropy)~\cite{skinnerMeasurementInducedPhaseTransitions2019} which employs the minimal cut formalism~\cite{nahumQuantumEntanglementGrowth2017}, and another for the $(n\geq 1)$-R\'{e}nyi entropies~\cite{jianMeasurementinducedCriticalityRandom2020,baoTheoryPhaseTransition2020} which uses the replica-trick to map the problem to 2D percolation in the limit of large local Hilbert space dimension $q \to \infty$. 

However, there is both analytical \cite{jianMeasurementinducedCriticalityRandom2020} and numerical evidence \cite{zabaloCriticalPropertiesMeasurementinduced2020,zabaloOperatorScalingDimensions2021,liConformalInvarianceQuantum2020} to suggest that away from this limit the universality class should be distinct from percolation. Puzzlingly, despite this evidence, numerics on 1+1D Haar-random and Clifford circuits give many \textit{bulk} exponents which are close to those of percolation. It has been suggested \cite{jianMeasurementinducedCriticalityRandom2020} that this could be an indication that the finite $q$ fixed point is close to the percolation fixed point in the RG phase diagram. 

Despite the results in 1+1D, it was not previously clear whether this proximity to percolation holds in higher dimensions. To address this, in the first part of our work we study the critical properties of the measurement-induced transition in 2+1D Clifford circuits. First, we precisely locate the critical point using the tripartite information $I_{3}$ (see \cref{sec:entanglement_transition}), which has been argued to be scale-invariant at criticality, thereby providing a good estimator of the critical probability $p_{c}$. Having fixed $p_{c}$, we then find an inverse power-law for the critical entanglement dynamics, $S(t,L) \sim L(1 - a/t)$, which saturates to an area-law (see Fig.~\ref{fig:summary}a). We provide a heuristic justification for this scaling based on the `minimal cut' prescription, which assumes a percolation-like picture. The steady-state area-law scaling is consistent with the behavior of conformal field theories in dimensions $d > 2$~\cite{calabreseEntanglementEntropyQuantum2004,fradkinEntanglementEntropy2D2006}.

We note that the accurate determination of the critical point using $I_{3}$ was important to correctly determine the critical scaling, since even small deviations can result in scaling which looks like $S \sim \mathcal{O}( L \log{L})$ (c.f.\ Ref.~\cite{turkeshiMeasurementinducedCriticalityDimensional2020} and the discussion in \cref{sec:cost_function}). 

Next we analyze the connection between this measurement-induced entanglement transition and quantum error-correction through the lens of the purification transition \cite{gullansDynamicalPurificationPhase2020,gullansScalableProbesMeasurementInduced2020,baoTheoryPhaseTransition2020,fanSelfOrganizedErrorCorrection2020,fidkowskiHowDynamicalQuantum2020}, which is characterized by a transition in the purification time of an initially maximally-mixed state---in the `mixed phase' the state purifies in a time exponential in system size $L$, whereas in the `pure phase' it purifies in a time polynomial in $L$. This purification transition can be viewed as a transition in the quantum channel capacity density of the hybrid quantum circuit, which governs whether the circuit can be used to generate a finite-rate quantum error-correcting code---the code rate is finite in the mixed phase, and goes to zero as one approaches the pure phase. In other words, these hybrid quantum circuits can form \textit{emergent} quantum error-correcting codes which protect against errors given precisely by the measurements involved in the circuit.

\begin{figure}[t]
    \centering
    \includegraphics[width=\columnwidth]{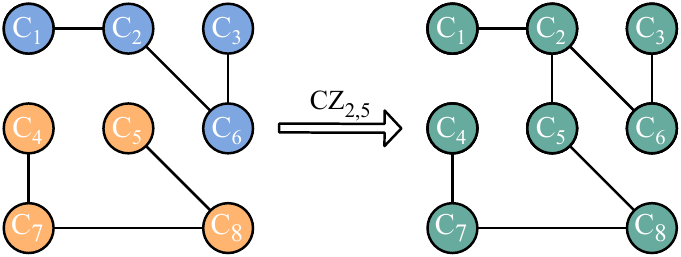}
    \caption{We employ a graph-state based simulation algorithm \cite{andersFastSimulationStabilizer2006}, where the data encoding the state consists of a graph $\mathcal{G}$ and a list $\{C_{i}\}_{i=1}^{L^{d}}$ of one-qubit Cliffords. The entanglement structure is completely fixed by $\mathcal{G}$. Entanglement clusters can be found by a breadth-first search on $\mathcal{G}$, and are here highlighted in different colors. In general the action of a Clifford gate corresponds to updating $\mathcal{G}$ and the list of one-qubit Cliffords. Here we illustrate the simple case of a CZ gate acting on two qubits whose one-qubit Cliffords commute with CZ; in this case the CZ gate simply toggles an edge between the qubits.}
    \label{fig:graph_algorithm_diagram}
\end{figure}

It is not \textit{a priori} obvious that these two measurement-induced transitions should coincide: the entanglement transition concerns spatial correlations in a quantum state at a fixed time, whereas the purification transition concerns correlations between quantum states at different times~\cite{gullansDynamicalPurificationPhase2020}. 
Their coincidence in 1+1D was explained by the fact that the 2+0D statistical mechanical model governing the purification transition is the same as that of the entanglement transition, just with different boundary conditions~\cite{baoTheoryPhaseTransition2020,liConformalInvarianceQuantum2020}. In these models, the time coordinate of the physical circuit plays the role of imaginary time in the stat-mech model, giving an emergent symmetry between space and time~\cite{liConformalInvarianceQuantum2020}.
In higher dimensions, however, the symmetry between space and time can be broken quite naturally. Our precise handle on the critical point allows us to demonstrate that the purification transition in 2+1D Clifford circuits continues to coincide with the entanglement transition, suggesting this phenomenon may be generic in all dimensions.

The coincidence of these two transitions then allows us to utilize the entangling and purifying dynamics of entangled ancilla qubits to extract various bulk and surface critical exponents of the transition in 2+1D Clifford circuits, and to provide evidence of conformal symmetry at the critical point (see \cref{sec:purification}). The \textit{bulk} exponents extracted in this way are within error-bars of 3D percolation (see \cref{tab:critical_exponents}). Interestingly, we do observe small deviations from percolation in certain surface critical exponents (see \cref{sec:purification}). This is similar to the behavior observed numerically in 1+1D circuits with qubits~\cite{gullansScalableProbesMeasurementInduced2020,zabaloCriticalPropertiesMeasurementinduced2020,liConformalInvarianceQuantum2020}. 

\begin{table}[t]
    \centering
    \begin{tabular}{lcccccc}
        \hline \hline
        & \multicolumn{2}{c}{Quantum circuits} &  & \multicolumn{3}{c}{Classical percolation} \\
        \cline{2-3} \cline{5-7}
        {\raisebox{1.5ex}[0pt]{Exponent}} & {1+1D C} & {2+1D C} &  & {1D P} & {2D P} & {3D P}\\
        \hline
        $\nu$ & 1.24(7) & 0.85(9) &  & 1 & $4 / 3 = 1.333$ & 0.8774\\
        $\eta$ & 0.22(1) & $-0.01(5)$ &  & 1 & $5 / 24 = 0.208$  & $- 0.047$\\
        $\eta_{\parallel}$ & 0.63(1) & \cellcolor{LightRed}0.85(4) &  & 1 & $2 / 3 = 0.667$ & 0.95\\
        $\eta_{\perp}$ & 0.43(2) & 0.46(8) &  & 1 & $7 / 16 = 0.438$ & 0.45\\
        $\beta$ & 0.14(1) & 0.40(1) &  & 0 & $5 / 36 = 0.139$ & 0.43\\
        $\beta_s$ & 0.39(2) & \cellcolor{LightRed}0.74(2) &  & 0 & $4 / 9 = 0.444$ & 0.85\\
        $z$ & 1.06(4) & 1.07(4) &  &  &  & \\
        \hline
        \multicolumn{7}{c}{Entanglement clusters} \\
        \hline
        $\beta_{ec} / \nu$ & $-0.009(2)$ & 0.00(2) & & & & \\
        $\beta_s / \nu$ &  &  &  & 0 & $1 / 3 = 0.333$ & 0.975\\
        $\beta / \nu$ & & &  & 0 & $5 / 48 = 0.104$ & 0.49\\
        $\gamma_{ec} / \nu$ & 0.95(1) & 1.84(2) & & & & \\
        $\gamma_{1,1} / \nu$ &  &  &  & 0 & $1 / 3 = 0.333$ & 0.049\\
        $\gamma / \nu$ & & &  & 1 & $43 / 24 = 1.792$ & 2.09\\
        $\tau$ & 2.04 & 1.98(1) &  & 2 & $187 / 91 = 2.055$ & 2.19\\
        \hline \hline
    \end{tabular}
    \caption{Critical exponents of the measurement-induced transition in hybrid 1+1D and 2+1D random Clifford circuits, compared with those of 1D, 2D and 3D percolation (1D P, 2D P, and 3D P respectively). Exponents which appear to differ from percolation are highlighted in red. Those exponents which describe the scaling of entanglement clusters are labelled by the subscript $ec$, and are compared with the bulk and surface exponents for percolation. The exponents for 1+1D Clifford circuits, excluding those describing entanglement clusters, are taken from Ref.~\cite{zabaloCriticalPropertiesMeasurementinduced2020}.}
    \label{tab:critical_exponents}
\end{table}

We perform our simulations using a graph-state based algorithm (see \cref{fig:graph_algorithm_diagram}) \cite{andersFastSimulationStabilizer2006}, which provides easy access to geometric information about the entanglement structure---the entanglement is completely fixed by the underyling graph. This allows us to employ graph-theoretic clustering tools to analyze \textit{entanglement clusters} in the steady-state (see \cref{sec:entanglement_clusters}). If we naively assume that the critical point has a simple geometric map to percolation, then the critical properties of these entanglement clusters should be governed by the \textit{surface} exponents of percolation, given that the clusters exist on the final timeslice of the $(d+1)$-dimensional bulk. To confirm this naive expectation, we first analyze entanglement clusters in the projective transverse field Ising model, which is a measurement-only Clifford model known to have a simple geometric map to percolation~\cite{langEntanglementTransitionProjective2020}. There we indeed find critical scaling of the entanglement clusters consist with surface percolation exponents.

However, moving on to the Clifford circuits, we find that, both in 1+1D and 2+1D, the entanglement clusters are governed by exponents significantly different from those of surface percolation (see Fig.~\ref{fig:summary}b). We interpret this as further evidence that the measurement-induced transition in qubit Clifford circuits is in a different universality class to percolation. Lessons from Haar-random circuits also tell us that, even when a map to percolation does exist, it may be highly non-trivial in nature, occurring for example only in a replica limit~\cite{jianMeasurementinducedCriticalityRandom2020,baoTheoryPhaseTransition2020}. The deviation from surface percolation exponents in the Clifford models indicates that, even if a map to percolation does exist in certain limits, it may not have such a simple geometric interpretation as do the analogous maps for the projective transverse field Ising model \cite{langEntanglementTransitionProjective2020} and the Hartley entropy in Haar-random circuits~\cite{skinnerMeasurementInducedPhaseTransitions2019}.

\section{Methods}

\subsection{Model}

\begin{figure}[t]
    \centering
    \includegraphics[width=\columnwidth]{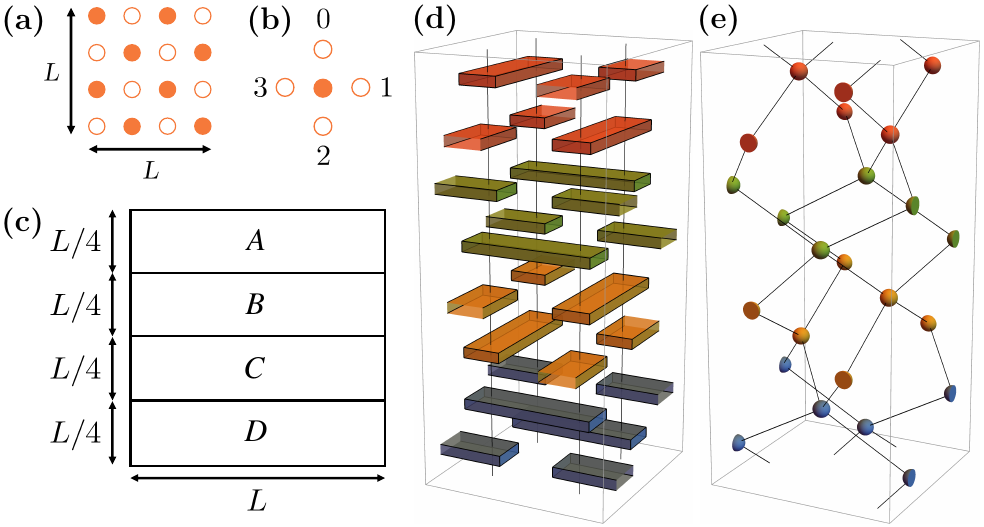}
    \caption{\textbf{(a)}~The sublattice index determines which sublattice of qubits, denoted by \filledDot or \unfilledDot, are used as the `controls' for the Clifford gates in that time step. \textbf{(b)}~Given a choice of sublattice index, the clock index determines in which direction each Clifford gate acts relative to the control. \textbf{(c)}~The geometry used to calculate the tripartite information $I_{3}(A : B : C)$. \textbf{(d)}~One period of the gate sequence on a $2 \times 2$ lattice with periodic boundary conditions, and time moving in the vertical direction. Different colors label different values of the clock index. \textbf{(e)}~Unit cell of the underlying lattice structure, obtained by contracting each Clifford gate into a point.}
    \label{fig:sublattice_diagram}
\end{figure}

In \cref{sec:entanglement_transition,sec:purification} we study a 2+1D model of local random Clifford dynamics interspersed with random projective measurements. Each time step consists of a round of random 2-qubit Clifford gates with disjoint support applied to nearest-neighbors, followed by a round of projective measurements in the $\sigma^{z}$ basis, where each qubit has probability $p$ of being measured. The gates are drawn uniformly over the whole 2-qubit Clifford group. The pattern of gates applied at a given time step is determined by two indices: a `sublattice index', which takes values in $\mathbb{Z}_{2}$, and a `clock index', which takes values in $\mathbb{Z}_{4}$. Arranging the qubits in an $L \times L$ square lattice with periodic boundary conditions, the sublattice index determines which sublattice of qubits will act as the `controls' for the Clifford gates (see \cref{fig:sublattice_diagram}a). Given a choice of sublattice, the clock index then determines which direction the Clifford gates act in relative to the control qubits. The values 0, 1, 2, and 3 correspond to gates acting up, right, down, and left from the control qubits respectively (see \cref{fig:sublattice_diagram}b). At the $n$th time step, the sublattice index has the value ${n \pmod{2}}$, and the clock index has the value $\lfloor n/2 \rfloor \pmod{4}$, so that the overall gate sequence has period 8 (see \cref{fig:sublattice_diagram}d). Since the support of the Clifford gates changes with each time step, certain quantities that depend on making a `cut', such as the entanglement entropy of a given region, exhibit a mild periodicity related to how often the gates cross the cut. To get well-defined steady-state values, we perform a window-average over a window matching the period of the oscillations (equal to 4 time steps in this case)---all quantities in this paper have been averaged in this way.

It is worth noting that this choice of gate protocol is by no means unique. On grounds of universality, we expect the main effect of a different choice of local quantum circuit is to change the critical measurement probability $p_{c}$, with the critical exponents unaffected. One alternative was explored in Ref.~\cite{turkeshiMeasurementinducedCriticalityDimensional2020}, which used 4-local gates instead of our 2-local gates. For rank-1 measurements, they observe a critical probability of $p_{c} \approx 0.54$, which is roughly the square root of our estimated value of $p_{c} \approx 0.312(2)$. They do observe a different correlation length exponent $\nu$, on which we comment in \cref{sec:entanglement_transition}. 

In \cref{sec:entanglement_clusters}, as well as studying the 2+1D Clifford model we have just outlined, we also study a 1+1D Clifford model. This is identical to that studied in many previous works studying the 1+1D problem, and can be thought of as being controlled by a single `sublattice index', resulting in a `brick-wall' structure of alternating layers of Clifford gates interspersed with random projective measurements.

\subsection{Simulation method}
\label{sec:method}

To simulate the hybrid Clifford dynamics, we used a graph-state based algorithm~\cite{andersFastSimulationStabilizer2006}. This makes use of the remarkable fact that every stabilizer state can be represented as a graph state, up to the action of some 1-qubit Cliffords~\cite{heinMultipartyEntanglementGraph2004}. Simulation of stabilizer states then takes the form of updating the underlying graph structure and the list of 1-qubit Cliffords, which can be done in polynomial time. 

In more detail, graph states are a class of pure quantum states whose structure is determined entirely by an underlying graph $\mathcal{G}=(V,E)$. Each graph vertex $v \in V$ corresponds to a qubit, and the graph edges $E$ determine the preparation procedure for the state. To prepare the graph state $|\mathcal{G}\rangle$, we start from the initial product state $| \psi_{0} \rangle = [(|0\rangle + |1\rangle)/\sqrt{2}]^{\otimes N}$, where $N$ is the number of qubits, and then apply a CZ gate to each pair of qubits which are connected by an edge in the graph $\mathcal{G}$. 

Stabilizer states are the states which can be prepared from the initial product state $|0 \rangle^{\otimes N}$ by acting with gates from the $N$-qubit Clifford group $\mathcal{C}_{N}$. The set of stabilizer states is larger than that of graph states, but not by much: all stabilizer states can be written as a graph state, up to the action of some gates from the 1-qubit Clifford group $\mathcal{C}_{1}$ [which contains only 24 gates, up to phase]. Single qubit gates are then trivial to perform, taking $\Theta(1)$ time. Two-qubit Cliffords take time $\mathcal{O}(d^{2})$, where $d$ is the maximum vertex degree of the qubits involved in the gate, and single-qubit Z-basis measurements take time $\mathcal{O}(d)$. This makes graphs with low connectivity, which can roughly be identified with low entangled states, easier to simulate.

To wit, the graph structure completely determines the entanglement of the corresponding quantum state. Given a bipartition of the system into subsystems $A$ and $B$, the (R\'{e}nyi or von Neumann) entanglement entropy $S_{A}$ is given by
\begin{equation}
    S_{A} = \mathrm{rank}(\Gamma_{AB}),
\end{equation}
where $\Gamma_{AB}$ is the submatrix of the adjacency matrix characterizing edges between subsystems $A$ and $B$~\cite{heinMultipartyEntanglementGraph2004}. We note that for stabilizer states all R\'{e}nyi entropies (including the von Neumann entropy) are equal~\cite{fattalEntanglementStabilizerFormalism2004}.

To simulate an initially mixed state $\rho$, we introduce an auxiliary system to obtain a purification of $\rho$. We then perform time-evolution on the resultant pure state, with the quantum circuit acting as the identity on the purifying system. For the maximally-mixed initial state on $N$ qubits, $\rho = \mathds{1} / 2^N$, this corresponds to the pure state simulation of $N$ Bell pairs, where the system dynamics acts only on one half of the Bell pairs. This purification simulation method does mean that $N$-qubit mixed states are harder to simulate than $N$-qubit pure states, but not as hard as $2N$-qubit pure states, since the purifying qubits typically have a lower vertex degree than the original qubits.

\subsection{Transition diagnostics}

As well as the entanglement entropy $S_{A}$, we also study the tripartite mutual information
\begin{equation}
    I_{3}(A : B : C) = I_{2}(A : B) + I_{2}(A : C) - I_{2}(A : BC),
\end{equation}
where $I_{2}(A : B) = S_{A} + S_{B} - S_{AB}$ is the mutual information. It is easy to see that for pure states, given a partition of the system into 4 subsystems, the tripartite information of 3 of the subsystems does not depend on the choice of subsystems, so from now on we will simply write $I_{3} \equiv I_{3}(A : B : C)$. We calculate $I_{3}$ for the partition shown in \cref{fig:sublattice_diagram}c. Notice that a vertical slice of this geometry gives a circle divided into four equal sections. In 1+1D this partitioning was successfully employed to study the entanglement transition because, at least within the minimal cut picture~\cite{nahumQuantumEntanglementGrowth2017}, it cancels out any boundary terms corresponding to the entanglement cost of a domain wall traversing from the circuit boundary to the percolating cluster in the bulk of the circuit~\cite{gullansDynamicalPurificationPhase2020,zabaloCriticalPropertiesMeasurementinduced2020}. This then suggests that in 1+1D, $I_{3}$ is extensive in the volume-law phase, $\mathcal{O}(1)$ at criticality, and zero in the area-law phase. In 2+1D, we argue that, for this particular choice of geometry, $I_{3}$ remains $\mathcal{O}(1)$ at criticality, with its overall behavior described by
\begin{equation}
    I_{3}(p,L) = \begin{cases}
                \mathcal{O}(L^{2}),& p < p_{c} \\
                \mathcal{O}(1),& p = p_{c} \\
                0,& p > p_{c}
            \end{cases}
    \label{eq:I3_scaling}
\end{equation}
This implies that the values of $I_{3}(p,L)$ should coincide for different system sizes at $p=p_{c}$, allowing for reliable location of the critical point. We further discuss our choice of geometry for $I_{3}$ in \cref{sec:cost_function}.

\section{Entanglement transition}
\label{sec:entanglement_transition}

\begin{figure}[t]
    \centering
    \includegraphics[width=\columnwidth]{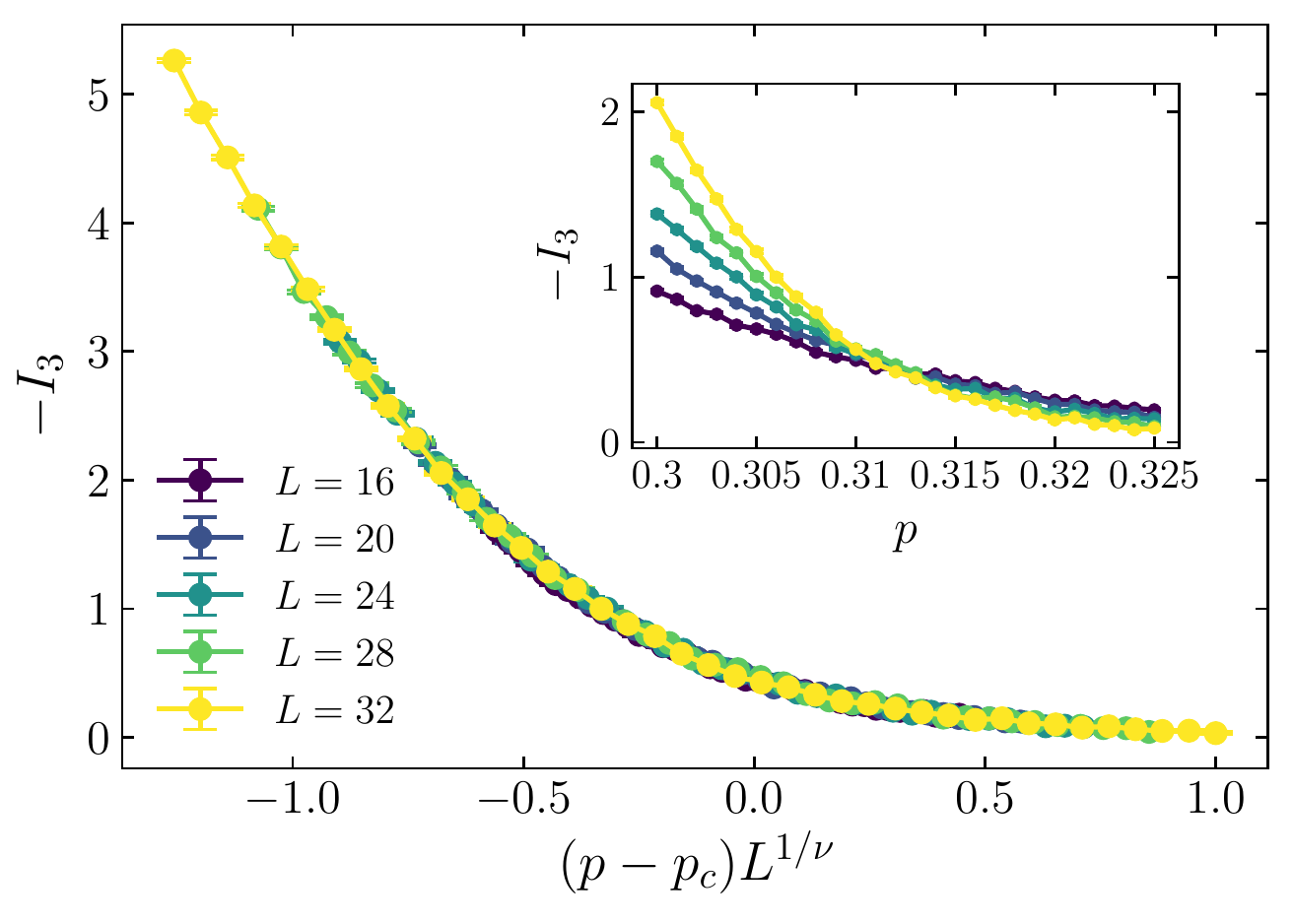}
    \caption{The steady-state $I_{3}$ as a function of $(p-p_{c})L^{1/\nu}$, where $p_{c} \approx 0.312(2)$ and $\nu \approx 0.85(9)$. The inset shows the uncollapsed data. This dataset consists of \SI{5e4} circuit realizations.}
    \label{fig:I3_transition}
\end{figure}

To accurately estimate the location of the critical point, it is necessary to determine the correct scaling of $I_{3}$. To that end, we must rule out plausible scalings which are different from the one proposed in \cref{eq:I3_scaling}. We have also investigated the possibility that $I_{3} \propto L$ at the critical point, which would suggest that the values of $I_{3}(p,L) / L$ should coincide at $p=p_{c}$. We detail evidence against this scaling form in \cref{sec:alternative_scaling}.

The steady-state values of $I_{3}(p,L)$ are plotted in \cref{fig:I3_transition}. Given the scaling in \cref{eq:I3_scaling}, the curves should coincide at the critical point. To determine the critical point and the correlation length exponent $\nu$ we make the finite-size scaling ansatz
\begin{equation}
    I_{3}(p,L) \sim F \left[(p-p_{c})L^{1/\nu}\right],
\end{equation}
where $F[\cdot]$ is a single-parameter scaling function. We determine the optimal parameters by minimizing a cost function $\epsilon(p_{c},\nu)$ which measures deviations of a point from a linear interpolation between its neighbors~\cite{kawashimaCriticalBehaviorThreeDimensional1993,zabaloCriticalPropertiesMeasurementinduced2020} (see \cref{sec:cost_function} for details). The resulting data collapse is of excellent quality, with $p_{c} \approx 0.312(2)$ and $\nu \approx 0.85(9)$, where the error bars correspond to the range of values for which the cost function is less than 2 times its minimum value. We note that this value of $\nu$ is reasonably close to the 3D percolation value of $\nu_{\mathrm{perc}} \approx 0.877$~\cite{kozaDiscreteContinuousPercolation2016}, suggesting that the close relationship between exponents of the entanglement transition and percolation, even at low local Hilbert space dimension, continues to hold in 2+1D. We also note that our value of $\nu$ is significantly larger than that reported in Ref.~\cite{turkeshiMeasurementinducedCriticalityDimensional2020} ($\nu \approx 0.67$); we attribute this to the fact that we extract $\nu$ by a data collapse not of the half-plane entanglement but of the tripartite information, which coincides for different system sizes at the critical point and so provides a much more accurate estimator of the critical point. A similar scenario occurs in 1+1D~\cite{zabaloCriticalPropertiesMeasurementinduced2020}. We discuss this further in \cref{sec:cost_function}.

Let us briefly comment on the value of $p_{c} \approx 0.312$ obtained for the critical measurement probability. This value coincides with the threshold for site percolation on the simple cubic lattice~\cite{xuSimultaneousAnalysisThreedimensional2014}, but as far as we are aware this is a coincidence; in fact our gate model maps to the lattice shown in \cref{fig:sublattice_diagram}e, which exhibits a bond percolation transition at $p_{c} = 0.3759(2)$. We expect other gate models to give different values of $p_{c}$ (see Ref.~\cite{turkeshiMeasurementinducedCriticalityDimensional2020}) but the same critical exponents. It is also interesting to compare our value of $p_{c}$ to the upper bound derived in Ref.~\cite{fanSelfOrganizedErrorCorrection2020}, which modeled the volume-law phase as forming a dynamically-generated non-degenerate quantum error-correcting code, allowing them to apply the quantum Hamming bound. The bound on $p_{c}$ depends only on the local Hilbert space dimension $q$ (not on the spatial dimension), and for $q=2$ gives $p_{c} \lesssim 0.1893$. While this bound was satisfied by 1+1D Haar-random and Clifford circuits ($p_{c} \approx 0.17$~\cite{zabaloCriticalPropertiesMeasurementinduced2020}), here we see that it is strongly violated in 2+1D Clifford circuits. A similar violation has also been observed in all-to-all models~\cite{gullansDynamicalPurificationPhase2020}, where it was pointed out that if these hybrid dynamics which violate this upper bound are to generate quantum error-correcting codes, these codes must be \textit{degenerate}. Finally, we note that the value of $I_{3}$ at criticality, $I_{3}^{2+1D}(p_{c}) = -0.47(8)$, is within error-bars of the value for 1+1D Clifford circuits, $I_{3}^{1+1D}(p_{c}) = -0.56(9)$~\cite{zabaloCriticalPropertiesMeasurementinduced2020}, suggesting the possibility that at criticality $I_{3}$ could reach an $\mathcal{O}(1)$ constant which is independent of dimension.

Having established the location of the critical point via finite-size scaling of $I_{3}$, we study the scaling properties of the entanglement entropy in the different phases. We propose the following scaling for the 2+1D circuit:
\begin{equation}
    S(p,L) \sim \begin{cases}
        L(1-\frac{a}{\xi}) + A \frac{L^2}{\xi^2},  & p < p_{c}, \\
        L,                                         & p = p_{c}, \\
        L(1-\frac{a}{\xi}),                        & p > p_{c},
    \end{cases}
    \label{eq:entropy_scaling}
\end{equation}
where $\xi = |p-p_c|^{-\nu}$ is the correlation length and $a, A$ are unknown constants. Such scaling implies the data collapse of the entropy is possible using a similar ansatz as in the 1+1D circuit~\cite{skinnerMeasurementInducedPhaseTransitions2019, liMeasurementdrivenEntanglementTransition2019},
\begin{equation}
    S(p,L) - S(p_c,L) = F[(p-p_c) L^{1/\nu}],
\end{equation}
where $F[\cdot]$ is a single-parameter scaling function, depending only on $L/\xi$.

\begin{figure}[t]
    \centering
    \includegraphics[clip,width=\columnwidth]{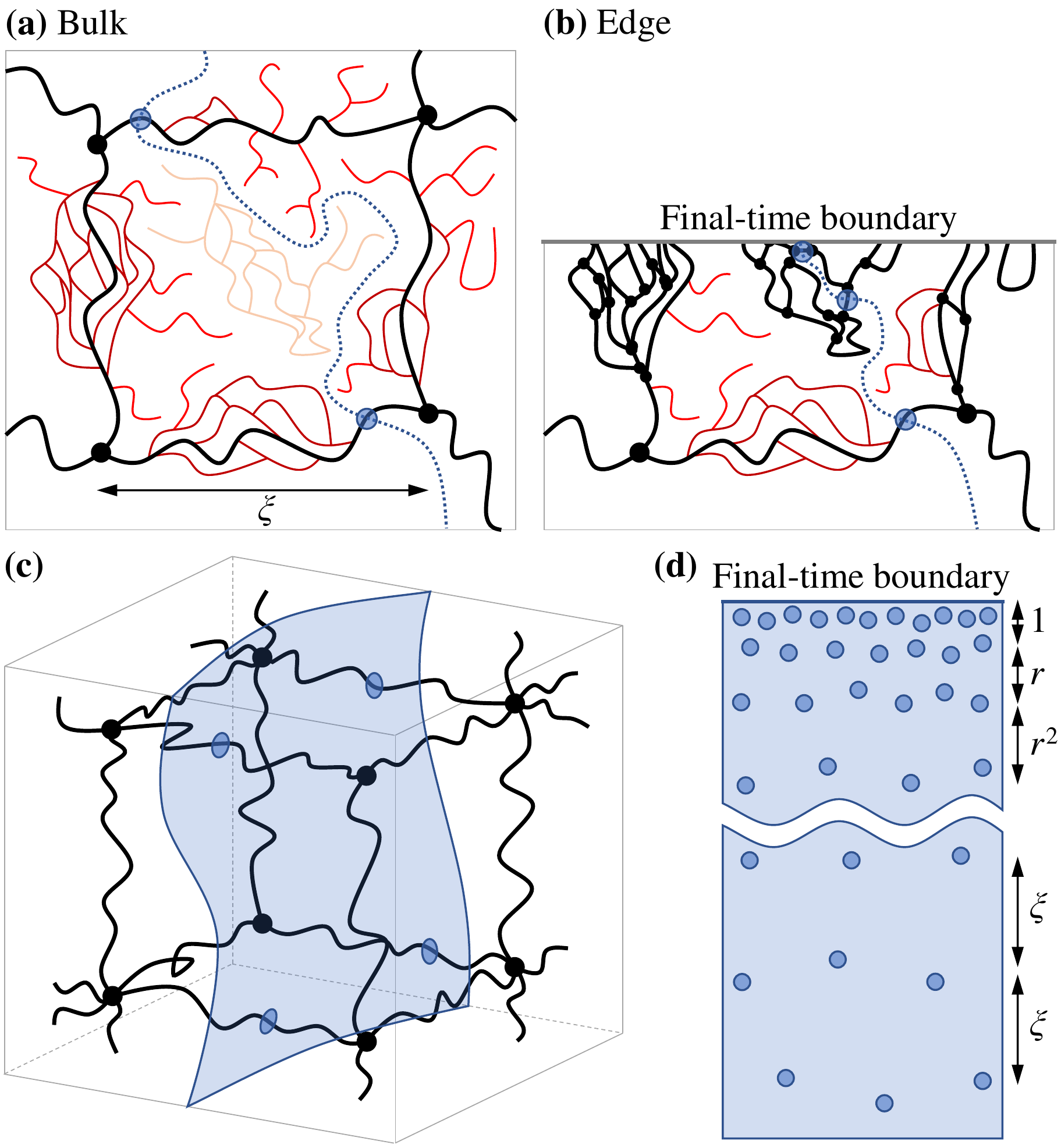}
    \caption{`Nodes and links' picture of percolation. \textbf{(a)}~An example of percolation in the bulk of a 2D system. Percolating bonds cluster within nodes (black dots) connected by links (thick black lines), forming a `wire frame'. Average distance between nodes is the correlation length $\xi$. There are also smaller structures on the links (dark red), dead ends (red) and structures unconnected to the frame (orange). Minimal-cut path (blue dotted line) can be deformed to only cut through the links (cuts indicated by transparent blue circles), causing an $O(1)$ contribution to the entropy. \textbf{(b)}~The same example, but in the presence of the final-time boundary. Every structure touching the edge is promoted to be part of the frame. Minimal-cut path generically starts within a smaller structure of size $O(1)$, having now to traverse through larger and larger chambers in order to reach structures of size $\xi$. \textbf{(c)}~Percolation in the bulk of a 3D system (showing only nodes and links for simplicity). Minimal-cut membrane can be deformed, contributing $O(1)$ to the entropy per one cell of the frame. \textbf{(d)}~Flattened minimal-cut membrane, showing all the necessary cuts. Near the edge, the membrane traverses layers of structures of increasingly larger sizes (with approximate common ratio $r$).}
    \label{fig:minimal_cut}
\end{figure}

In order to see the origin of this proposed scaling form, we draw from the similarity to the 1+1D case, where the behavior of entropy can be intuitively understood by considering the Hartley entropy $S_0$. For Haar random circuits $S_0$ can be mapped exactly to classical percolation in 2D~\cite{skinnerMeasurementInducedPhaseTransitions2019}: each projective measurement cuts a bond of the underlying lattice and prevents percolation; Hartley entropy of a region is then calculated as the minimal number of cuts needed to separate said region at the final-time boundary from the rest of the circuit. This mapping extends naturally to $d$+1D circuits, where $S_0$ corresponds to a minimal-cut $d$-dimensional membrane. Near criticality, the `nodes-and-links' picture of percolation~\cite{skalTopologyInfiniteCluster1975, staufferIntroductionPercolationTheory2018} gives an insight into the scaling properties of $S_0$ (see \cref{fig:minimal_cut}) and shows two important contributions: from the bulk, and from the edge. 

\begin{figure}[t]
    \centering
    \includegraphics[width=\columnwidth]{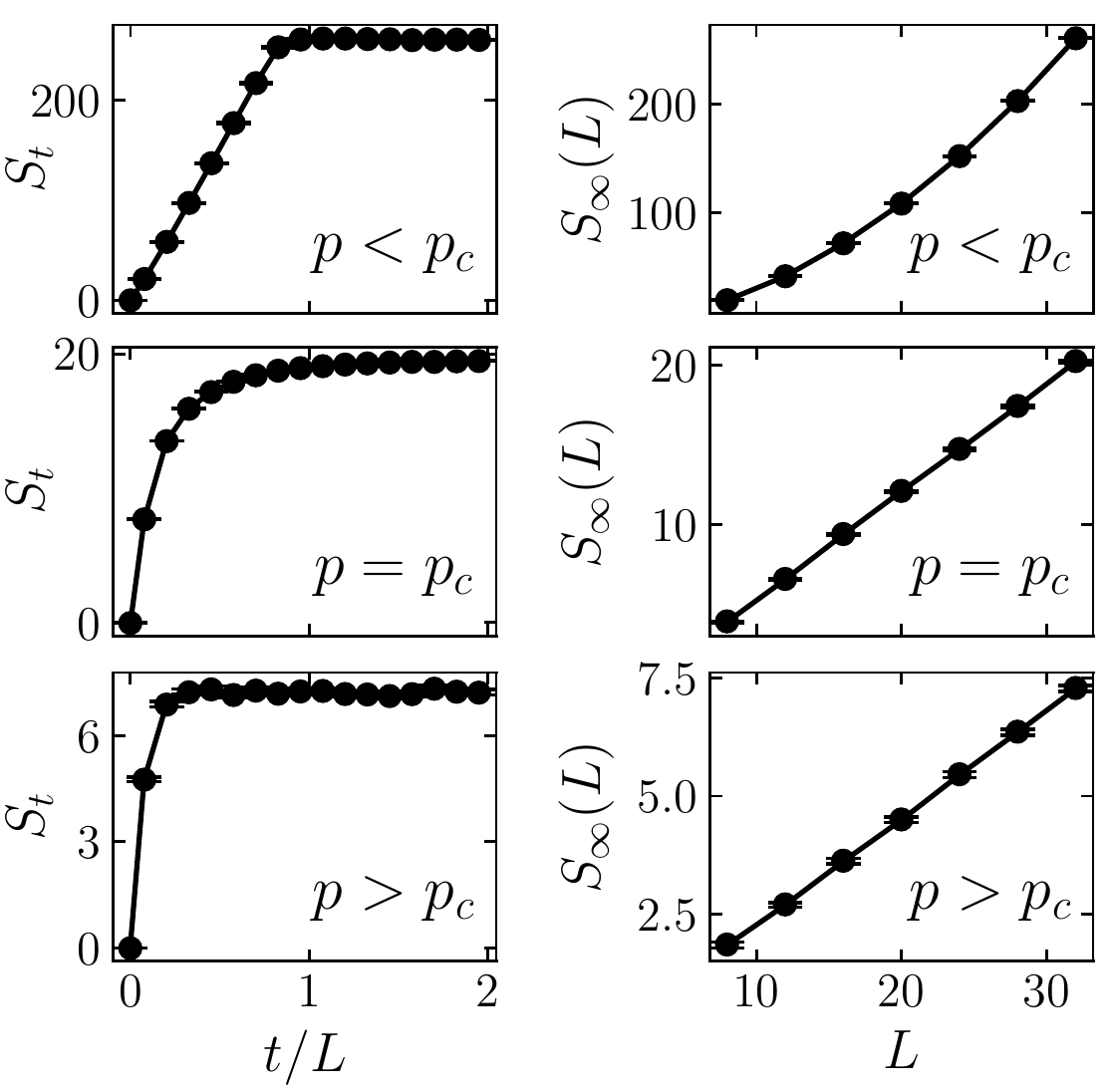}
    \caption{Dynamics and steady-state behavior of the half-plane entanglement $S(L/2 \times L)$ in the volume-law ($p<p_{c}$), critical ($p=p_{c}$), and area-law ($p>p_{c}$) phases. The left column shows the dynamics for $L=32$, with $S_{t} \sim L t$ for $p<p_{c}$, $S_{t} \sim L(1 - a/t)$ for $p=p_{c}$, and $S_{t}$ saturating in $\mathcal{O}(1)$ time for $p>p_{c}$. The right column shows the steady-state scaling, with $S_{\infty}(L) \sim \mathcal{O}(L^{2})$ for $p<p_{c}$, and $S_{\infty}(L) \sim \mathcal{O}(L)$ for $p\geq p_{c}$. We use $p=0.1$, $p=0.312$, and $p=0.4$ for the volume-law, critical, and area-law plots respectively.}
    \label{fig:entropy_scaling}
\end{figure}

For $p<p_c$, percolation in the bulk of the circuit is possible due to unbroken bonds forming a `wire frame' consisting of dense clusters of bonds (nodes) connected by long chains of unbroken bonds (links). Each cell in the frame is of the size of the correlation length $\xi$ and, if traversed by the minimal-cut membrane, gives a contribution of $O(1)$ to the entropy [see \cref{fig:minimal_cut}(a) and (c) for 2D and 3D examples]. Counting the number of cells results in the bulk of the circuit contributing $\sim (L/\xi)^d$ to $S_0$, the source of the volume-law scaling.

The second relevant contribution comes from the final-time boundary of the circuit [see \cref{fig:minimal_cut}(b)]. This edge cuts through not only the links and nodes discussed above, but also through smaller structures, dead ends, and other structures normally unconnected to the main mesh. This results in the minimal-cut membrane having to generically cut through a large number of small mesh cells right next to the boundary, then through layers of consecutively larger cells, until the cell size reaches $\xi$ [see \cref{fig:minimal_cut}(d)]. Assuming a geometric progression of cell sizes with common ratio $r>1$~\cite{skinnerMeasurementInducedPhaseTransitions2019}, the number of cells in the $i$th layer is $\sim (L/r^i)^{d-1}$, while the total number of layers is $\sim \log_r \xi$. We then arrive at an important result: the total contribution from the boundary for 1+1D is $\sim \log \xi$, while for higher dimensions is $\sim (1 - a / \xi^{d-1}) L^{d-1}$. This term is in general responsible for the area-law scaling, but at the critical point $p=p_c$ (when $\xi\to L$) it results in \textit{logarithmic} scaling in 1+1D, and \textit{area-law} scaling in higher dimensions.

We can also use this analysis to predict the time-dependence of the entanglement entropy at criticality. For intermediate times $1 \ll t \ll \min(\xi,L)$, the circuit is shallow, and the minimal cut membrane will pass from the final time boundary to the initial time boundary. This is because at $t=0$ the system is in a product state and the membrane can traverse the initial boundary freely. Hence, the main contribution to the entropy will be from summing over progressively larger cells up until the circuit depth of $t$, i.e. the number of layers is now only $\sim \log_{r}{t}$. Thus, the geometric sum $\sum_{i}^{\log_r t} (L/r^i)^{d-1}$ gives
\begin{equation}
    S(t,L) \sim L^{d-1} \left( 1 - \dfrac{a}{t^{d-1}}\right)
    \label{eq:entropy_dynamics}
\end{equation}
for some $\mathcal{O}(1)$ constant $a$. For the special case of $d = 1$ the sum reduces to the logarithmic scaling $S(t,L) \sim \log{t}$~\cite{skinnerMeasurementInducedPhaseTransitions2019}, but in higher dimensions the growth takes the form of an inverse power-law in time, eventually saturating to an area-law. We can write this as a scaling form $S(t,L) - b L^{d-1} \sim f(t/L)$ with $f(x) \sim - x^{-(d-1)}$ as $x \to 0$ and $f(x) \to \mathrm{const.}$ as $x\to \infty$, consistent with a dynamical critical exponent of $z=1$ (see also \cref{fig:purification_transition}b and \cref{fig:critical_entanglement_dynamics}c).

\cref{fig:entropy_scaling} presents a summary of our results for the entanglement entropy, showing an excellent agreement with the scaling ansatze in \cref{eq:entropy_scaling,eq:entropy_dynamics}. Notably, in the steady state we observe \textit{area-law} scaling at the critical point (consistent with the recent results of Ref.~\cite{lavasaniTopologicalOrderCriticality2020}), possibly with subleading \textit{additive} logarithmic corrections, but \textit{not} with multiplicative logarithmic corrections ($L \log L$), as implied in Ref.~\cite{turkeshiMeasurementinducedCriticalityDimensional2020}. We note however that if one assumes a lower transition point ($p\approx 0.29$), numerics may seem like a $L \log L$ behavior for small system sizes, suggesting that correctly locating the critical value $p_c$ is crucial to making any statements on scaling of entropy at criticality. As explained above, data collapse of $I_3$ pinpoints the precise value of $p_c$, allowing us to determine the correct critical scaling behavior.

Moreover, at these system sizes we cannot directly observe the presence of a subleading additive $\log{L}$ term, but we also cannot rule it out since it may have a small coefficient. Such a subleading additive $\log{L}$ is predicted by a calculation from capillary wave theory~\cite{gelfandFinitesizeEffectsFluid1990,liStatisticalMechanicsQuantum2020} which evaluates the free energy cost of inserting an Ising domain wall membrane in the quantum circuit's spacetime bulk, with the boundary condition that at the boundary of the circuit corresponding to the final time the membrane is pinned to the region for which one wants to calculate the entanglement entropy. The subleading $\log{L}$ then corresponds to an entropic contribution to the free energy from `thermal' fluctuations of the membrane at finite `temperature' (here corresponding to nonzero measurement probability). In general, the appearance at criticality of an area-law with additive log corrections is reminiscent of the behavior of higher-dimensional conformal field theories~\cite{calabreseEntanglementEntropyQuantum2004,fradkinEntanglementEntropy2D2006}. There is also the possibility of a sublinear power-law correction, analogous to the $\sim L^{0.38}$ correction observed numerically in 1+1D Clifford circuits~\cite{liStatisticalMechanicsQuantum2020}, which could indicate a more complex entanglement domain wall structure than the simple Ising structure that predicts the logarithmic correction.

Finally, regarding the critical entanglement dynamics, we note that one must be careful to distinguish the inverse power-law behaviour of \cref{eq:entropy_dynamics} from logarithmic growth. In \cref{sec:appendix_entanglement_dynamics} we provide a plot of the critical entropy dynamics at $L=92$ on a log scale, which demonstrates that the growth is not logarithmic in time, and provide further evidence for the inverse power-law scaling.

\section{Purification transition}
\label{sec:purification}

\begin{figure}[t]
    \centering
    \includegraphics[width=\columnwidth]{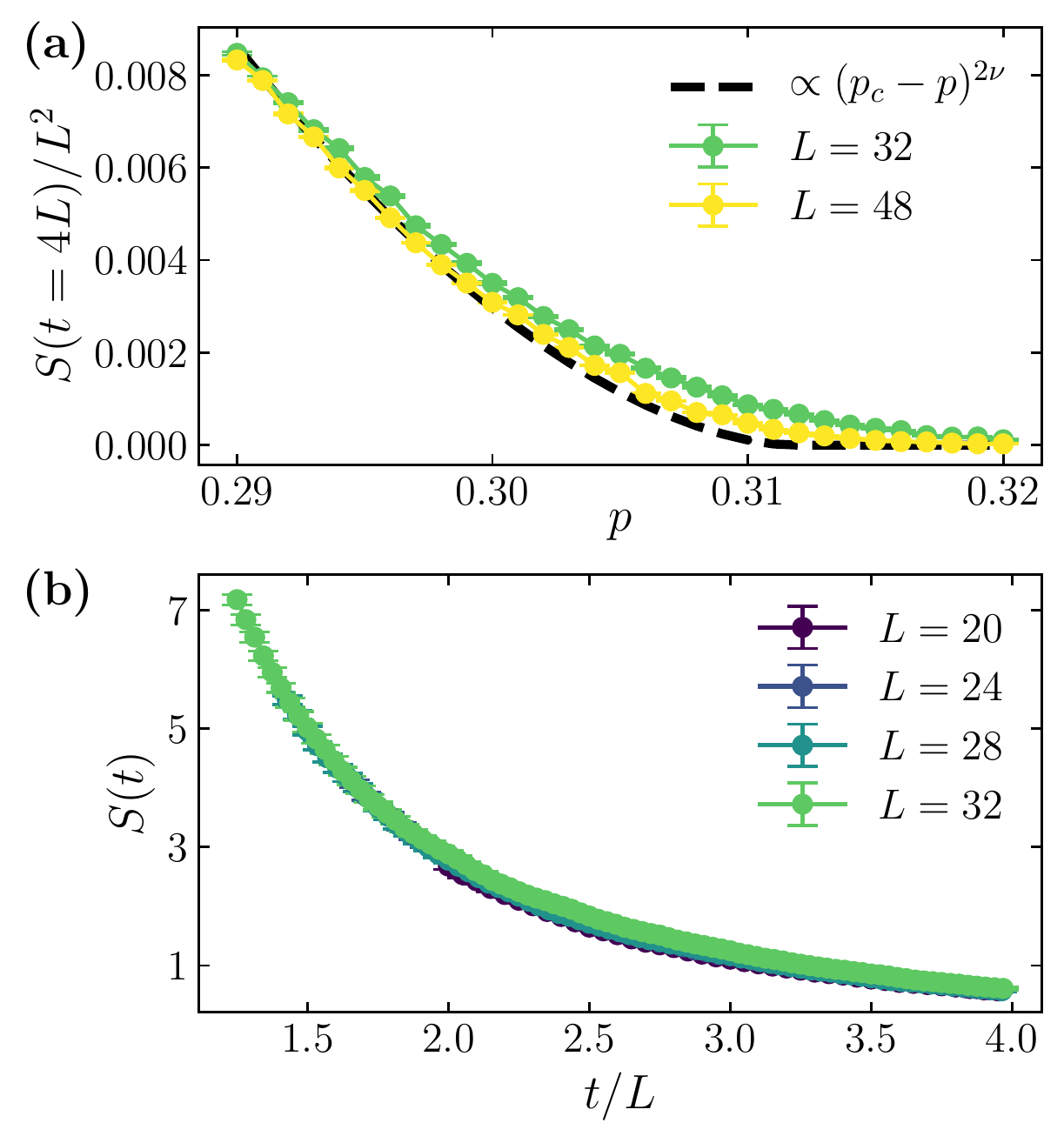}
    \caption{\textbf{(a)} The entropy density of an initially maximally-mixed state after evolving for a time $t=4L$. The black dashed line shows the function $A(p_{c} - p)^{2\nu}$ with $A \approx 11.7$, and $p_{c}$ and $\nu$ determined from finite-size scaling of $I_{3}$. At these system sizes there is still some finite-size drift in the data, but it seems to be approaching the curve described by $A(p_{c} - p)^{2\nu}$. \textbf{(b)} Purification dynamics at $p=p_{c}$. The data collapse onto a single curve when plotted in terms of $t/L$, indicating a dynamical critical exponent of $z\approx 1$ [the optimal fitted value is $z = 1.07(4)$]. Non-universal early-time dynamics are excluded from the fit.}
    \label{fig:purification_transition}
\end{figure}

In this section, we investigate the purification transition and demonstrate that it coincides with the entanglement transition studied in \cref{sec:entanglement_transition}. To do so we study the entanglement entropy density $S/L^{2}$ of a maximally-mixed initial state after being time-evolved for time $t=4L$. In the `pure phase', the state purifies in time linear in system size $L$, implying $S/L^{2} \to 0$ for $t \propto L$ but sufficiently large ($t=4L$ suffices), while in the `mixed phase' the purification time is exponential in $L$, so that after the time $t=4L$ we expect the entropy density to remain finite. \cref{fig:purification_transition} shows the entanglement entropy density as a function of measurement probability $p$. The entropy density vanishes close to the critical point $p_{c} \approx 0.312$ of the entanglement transition. For these system sizes, there still exists some appreciable finite-size drift, but it appears to be such that the entropy density vanishes increasingly close to $p_{c} \approx 0.312$ as the system size increases. The black dashed curve shows the function $A(p_{c}-p)^{2\nu}$, with $A$ a constant and $p_{c}$ and $\nu$ fixed from the entanglement transition. The exponent $2\nu$ is motivated by the scaling of the entanglement entropy in \cref{eq:entropy_scaling}, where the $\mathcal{O}(L^{2})$ term controlling the entropy density appears with the coefficient $\xi^{-2} \sim (p_{c} - p)^{2\nu}$. The convergence of the entropy density to the scaling form $A(p_{c}-p)^{2\nu}$ therefore provides strong evidence that the purification transition indeed coincides with the entanglement transition and that the estimation of $\nu$ in the previous section is correct.

\begin{figure}[t]
    \centering
    \includegraphics[width=\columnwidth]{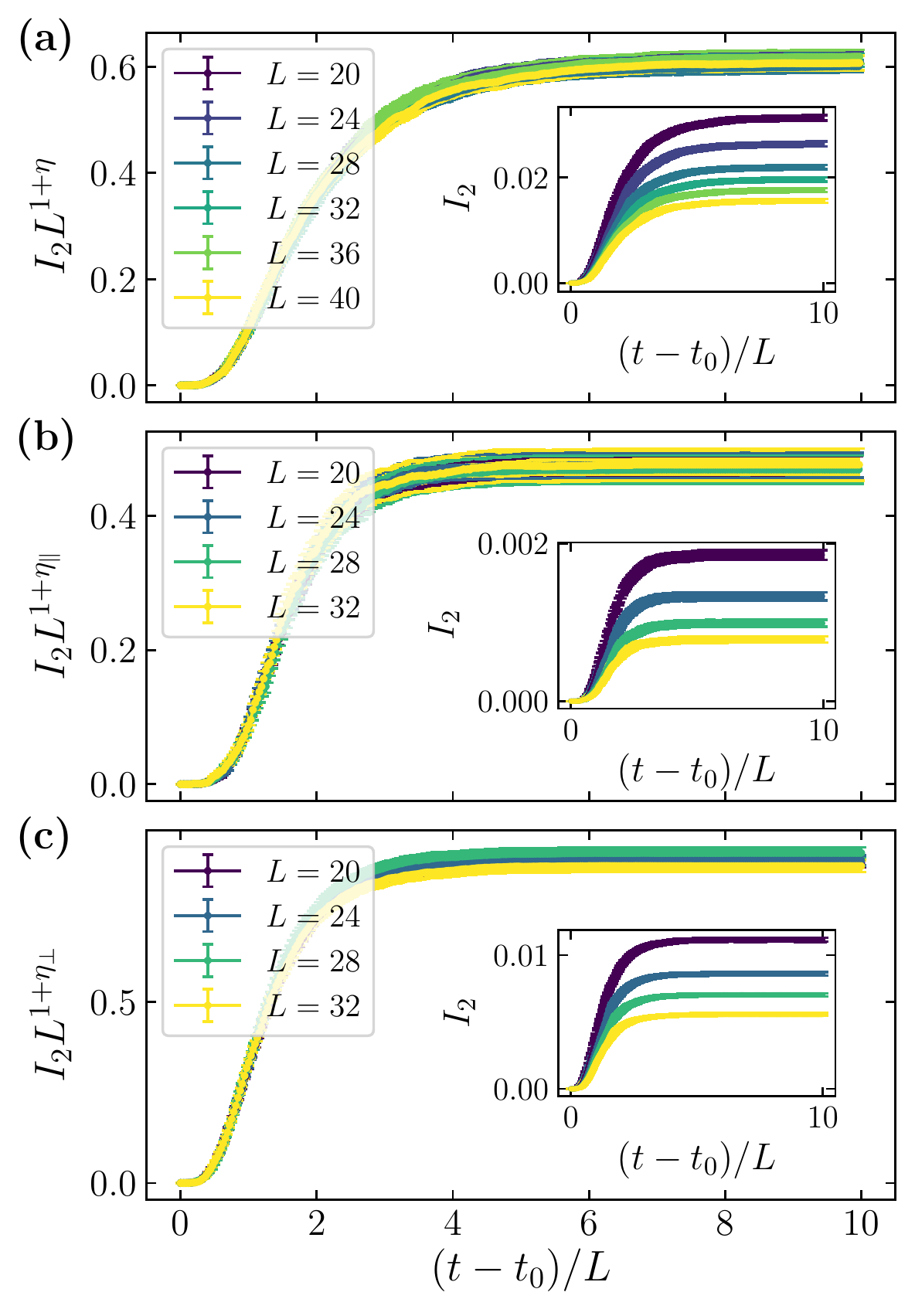}
    \caption{Extraction of the anomalous scaling exponents $\eta \approx -0.01(5)$, $\eta_{\parallel}\approx 0.85(4)$, and $\eta_{\bot} \approx 0.46(8)$, shown in \textbf{(a)}, \textbf{(b)}, and \textbf{(c)} respectively, via data collapse at $p=p_{c}$ of the mutual information $I_{2}$ between two ancilla qubits which are entangled at time $t_{0}$ with two system qubits a distance $L/2$ apart. The different exponents are extracted using different boundary conditions and different values of $t_{0}$ [see main text]. The insets show the uncollapsed data. The $\eta$ dataset consists of \SI{2.5e5} circuit realizations, while the $\eta_{\parallel}$ and $\eta_{\bot}$ datasets each consist of \SI{e6} circuit realizations.}
    \label{fig:eta_exponent}
\end{figure}

\begin{figure}[t]
    \centering
    \includegraphics[width=\columnwidth]{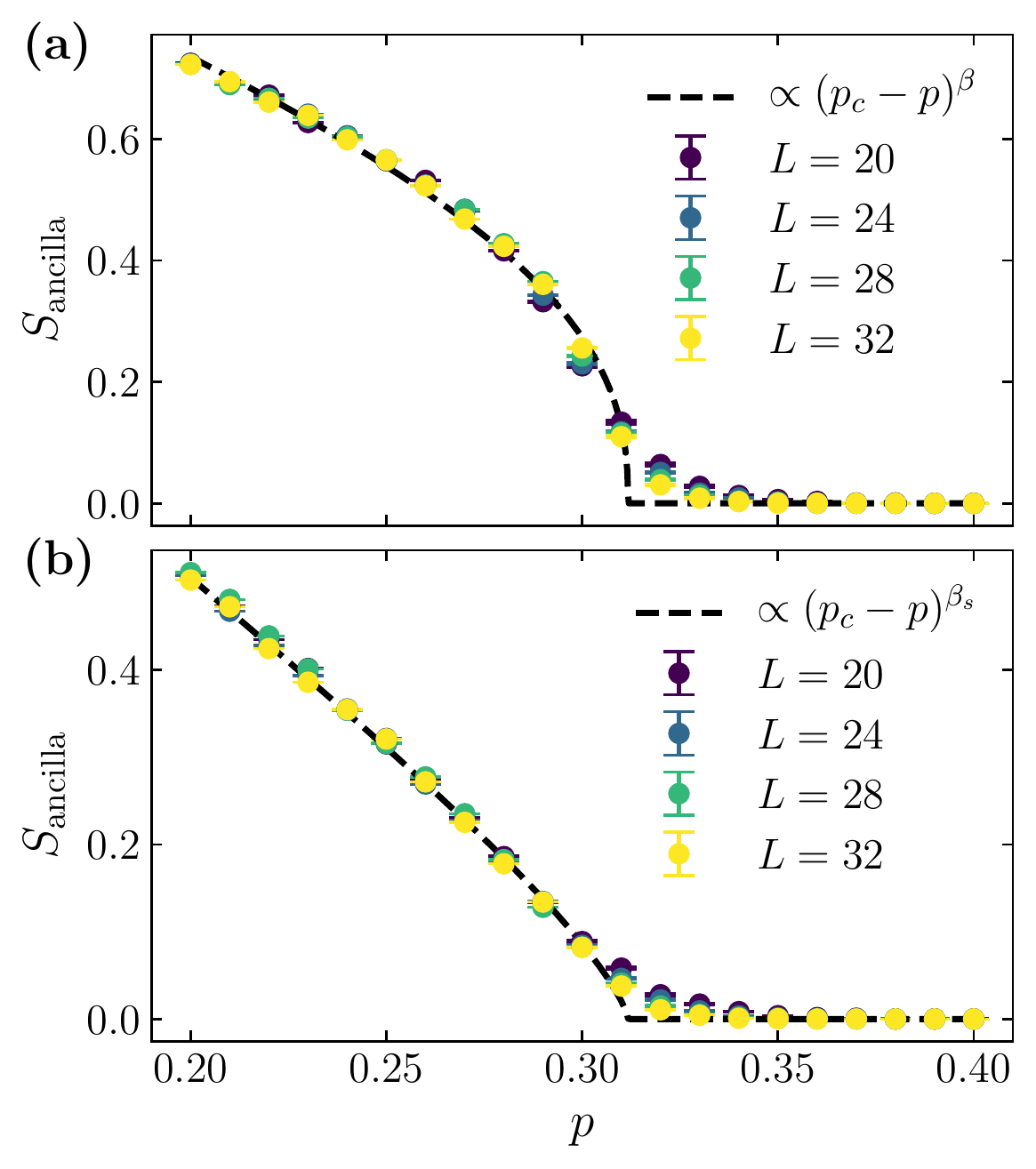}
    \caption{Extracting the exponents $\beta$ and $\beta_{s}$ using the entropy $S_{\mathrm{ancilla}}$ of an ancilla qubit which is maximally entangled with a bulk qubit at a time $t_{0}$, and then further evolved for a time $t=2L$. \textbf{(a)} The bulk exponent $\beta$ is extracted using $t_{0} = 2L$. The black dashed curve shows the function $B(p_{c} - p)^{\beta}$ where $B \approx 3.2$ and $\beta \approx 0.40(1)$. \textbf{(b)} The surface exponent $\beta_{s}$ is extracted using $t_{0} = 0$. There the black dashed curve shows the function $C(p_{c} - p)^{\beta_{s}}$ where $C \approx 4.6$ and $\beta_{s} \approx 0.74(2)$. In both cases, $p_{c} \approx 0.312$ is fixed by finite-size scaling of $I_{3}$. This dataset consists of \SI{e4} circuit realizations.}
    \label{fig:beta_critical_exponents}
\end{figure}

Having established the coincidence of these two transitions, we now extract further critical exponents of the transition using the local order parameter proposed in Ref.~\cite{gullansScalableProbesMeasurementInduced2020} of the entanglement entropy of an ancilla qubit entangled with the system but not directly acted on by the circuit dynamics. First, we extract the anomalous scaling exponents $\eta$, $\eta_{\parallel}$, and $\eta_{\bot}$ controlling the power-law decay of bulk-bulk, surface-surface, and surface-bulk two-point correlation functions at criticality. In percolation, these quantities control the probabilities that two distant sites, living either in the bulk or on the surface, belong to the same cluster. To determine these exponents we study the dynamics at $p=p_{c}$ of the mutual information between two ancilla qubits separated by a distance $L/2$~\cite{zabaloCriticalPropertiesMeasurementinduced2020}, which provides an upper bound on connected correlation functions~\cite{wolfAreaLawsQuantum2008}. The ancilla qubits are entangled with the system at a time $t_{0}$. We use different values of $t_{0}$ and different boundary conditions to extract the different exponents: \{$t_{0}=2L$, periodic\} for $\eta$, \{$t_{0}=0$, periodic\} for $\eta_{\parallel}$, and \{$t_{0}=2L$, open\} for $\eta_{\bot}$. Conformal symmetry $z=1$ at the critical point (see \cref{fig:purification_transition}b) implies that in $D$ spacetime dimensions the mutual information between two qubits separated by a distance $r$ should assume the scaling form
\begin{equation}
    I_{2}(t,r) \sim \dfrac{1}{r^{D-2+\eta}}\, G\left[\dfrac{t-t_{0}}{r}\right], 
\end{equation}
where $G[\cdot]$ is a single-parameter scaling function, and the exponent depends on the choice of $t_{0}$ and boundary conditions, as outlined above. Thus in this 2+1D spacetime circuit, we can extract the exponents by performing data collapses of $L^{1+\eta} I_{2}[(t-t_{0})/L,L/2]$, as shown in \cref{fig:eta_exponent}. For the bulk-bulk exponent $\eta$ and the surface-bulk exponent $\eta_{\bot}$, we obtain the values $\eta \approx -0.01(5)$ and $\eta_{\bot} \approx 0.46(8)$, which are within error-bars of the 3D percolation values $\eta_{\mathrm{perc}} = -0.047$ and $\eta_{\bot,\mathrm{perc}} = 0.45$~\cite{graceyFourLoopRenormalization2015}. We note in passing that the data collapse for $\eta_{\bot}$ is not as good quality as that for $\eta$, resulting in larger error bars using the methodology described in \cref{sec:cost_function}. However, there does not appear to be a systematic drift with increasing system size. We attempted to improve the collapse quality by using a large number of circuit realizations ($10^{6}$ for $\eta_{\bot}$), but some discrepancy is still evident. This could possibly be a result of $\eta_{\bot}$ being particularly sensitive to any miscalibration of the critical point $p_{c}$, despite the precision to which we have pinpointed $p_{c}$ in this work.

Moving on to the surface-surface exponent $\eta_{\parallel}$, we obtain the value $\eta_{\parallel} \approx 0.85(4)$. This is \textit{not} within error-bars of the 3D percolation value $\eta_{\parallel,\mathrm{perc}} = 0.95$, indicating a possible difference in surface behavior. The error-bars on our exponent estimates capture only the statistical error, so it is possible that there are still significant finite-size corrections. However, we note that a similar deviation in $\eta_{\parallel}$ (and in $\eta_{\bot}$), was observed in 1+1D Haar-random circuits (though not in Clifford circuits)~\cite{zabaloCriticalPropertiesMeasurementinduced2020}. In this case, a deviation only in $\eta_{\parallel}$ would not be consistent with the scaling relation $2\eta_{\bot} = \eta + \eta_{\parallel}$, but the error-bars on our estimates are large enough that there could also be small deviations in $\eta_{\bot}$ that provide the necessary contribution to restore the scaling relation.

Next, we extract the exponents $\beta$ and $\beta_{s}$ controlling the behavior of the order parameter as a function of $p$. In percolation, $\beta$ controls the probability $P(p) \sim |p - p_{c}|^{\beta}$ that a site in the bulk will belong to the infinite percolating cluster, while $\beta_{s}$ does the same but for a site on the surface. To extract these exponents we study the entanglement entropy of an ancilla qubit, entangled with the system at time $t_{0} = 2L$ for $\beta$ and time $t_{0} = 0$ for $\beta_{s}$, and subsequently time-evolved for a further time $t=2L$. \cref{fig:beta_critical_exponents} shows the ancilla entropy $S_{\mathrm{ancilla}}$ as a function of measurement probability $p$ for the cases relevant to $\beta$ and $\beta_{s}$. For the bulk exponent $\beta$, the data are well described by the function $B(p-p_{c})^{\beta}$ with $B$ a constant, $p_{c} \approx 0.312$ fixed by the entanglement transition, and $\beta \approx 0.40(1)$. This is close to the 3D percolation value of $\beta_{\mathrm{perc}} \approx 0.43$. However, for the surface exponent $\beta_{s}$, the data are well described by the function $C(p_{c} - p)^{\beta_{s}}$, where $\beta_{s} \approx 0.74(2)$. This is somewhat different from the 3D percolation value of $\beta_{s,\mathrm{perc}} \approx 0.85$. The value of $\beta_{s}$ is quite sensitive to the value of $p_{c}$; we estimate that to obtain $\beta_{s} \approx 0.85$ one would have to have $p_{c} \approx 0.318$, which does not seem tenable given the clear crossing point in $I_{3}$ (see inset of \cref{fig:I3_transition}). There are also some small deviations from the scaling around $p \approx p_{c}$, but these seem to decrease with system size. We therefore tentatively conclude that the surface critical exponent $\beta_{s}$ may also differ from 3D percolation. The fact that we observe both the surface exponents $\beta_{s}$ and $\eta_{\parallel}$ to be smaller than the corresponding values from percolation is consistent with the scaling relation $2\beta_{s} = \nu(D - 2 + \eta_{\parallel})$, where $D = 3$ is the number of spacetime dimensions.

\section{Entanglement clusters}
\label{sec:entanglement_clusters}

Finally, with the aim of further exploring connections with percolation, we investigate entanglement clusters in the steady state. Working within the graph-state framework for simulating stabilizer states, we define an entanglement cluster in the graph-theoretic sense: two spins are in the same cluster if there is a connected path between them (see \cref{fig:graph_algorithm_diagram} for an example). We will mainly study the size $s$ of the clusters, defined for a given cluster as the number of spins it contains. This is clearly quite a coarse-grained notion of entanglement, since different spins in the same cluster can be entangled by different amounts. Nonetheless, it provides some insight into how multipartite is the steady-state entanglement. 

\begin{figure}[t]
    \centering
    \includegraphics[width=\columnwidth]{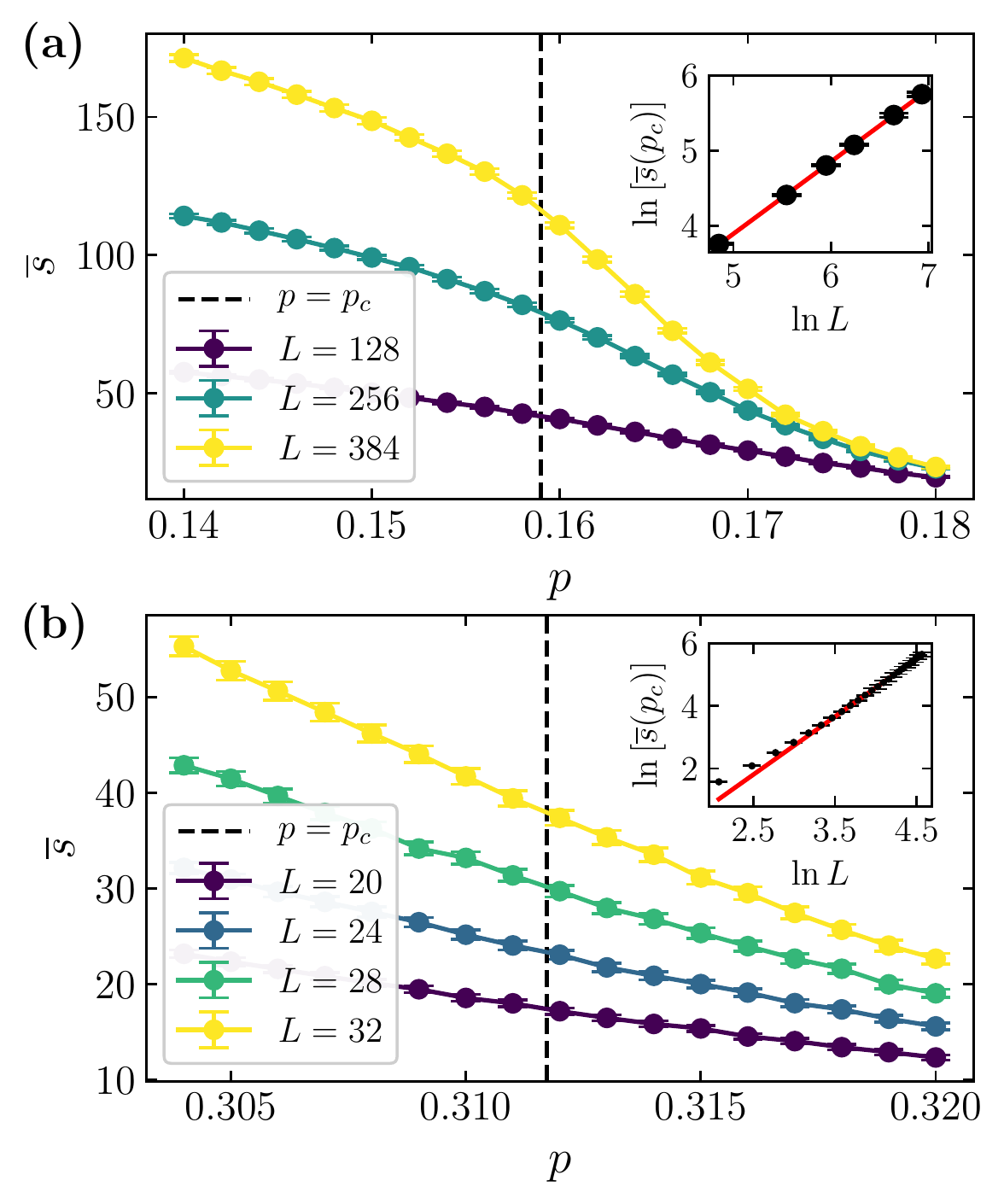}
    \caption{The average size $\overline{s}$ of all entanglement clusters in the steady-state for \textbf{(a)} 1+1D and \textbf{(b)} 2+1D Clifford circuits. The insets show log-log plots of this quantity at $p=p_{c}$, with the behaviour well described by the power law $\overline{s} \sim L^{\gamma_{ec}/\nu}$, where $\gamma_{ec}/\nu = 0.95(1)$ for 1+1D and $\gamma_{ec}/\nu = 1.84(2)$ for 2+1D (power-law fits shown in solid red).}
    \label{fig:mean_cluster_size}
\end{figure}

If we assume that there is a percolation-like statistical mechanical model controlling the critical point, then naively one would expect the scaling of the entanglement clusters to be controlled by surface exponents of $(d+1)$-dimensional percolation. We will focus on two quantities, the largest entanglement cluster size $s_{\mathrm{max}}$, and the mean entanglement cluster size $\overline{s}$. Within the percolation language, these correspond to the `surface area' of the infinite percolating cluster (assuming the largest surface cluster coincides with the largest bulk cluster), and the mean `surface area' of clusters with at least one site on the surface, where by `surface area' we mean the number of sites in the cluster that lie on the surface. In a $(d+1)$-dimensional percolation model with finite linear extent $L$, these should scale as $s_{\mathrm{max}} / L^{d} \sim L^{-\beta_{s}/\nu}$ and $\overline{s} \sim L^{\gamma_{1,1} / \nu}$ respectively.

To check this naive expectation, we first analyzed the scaling of entanglement clusters within the projective transverse field Ising model (PTFIM), as discussed in more detail in \cref{sec:PTFIM_entanglement_clusters}. This is a measurement-only model exhibiting an entanglement transition which is known to be in the percolation universality class \cite{langEntanglementTransitionProjective2020}. Conveniently, it also only involves Clifford operations, so can be simulated using the graph-state framework, and therefore provides a useful testbed for the scaling properties of the entanglement clusters. The results are summarized in \cref{fig:projective_TFIM_combined_plots}, where we show data for the mean cluster size and largest cluster size for the PTFIM in both 1+1D and 2+1D. In 1+1D, these quantities both scale as power-laws with exponents closely matching the expected values from surface 2D percolation. In 2+1D, the largest cluster size also follows a power-law closely matching the expectation from surface 3D percolation. The mean cluster size appears to have a slightly larger exponent than expected, but it is possible that this discrepancy is due to significant finite-size effects, as we discuss in more detail in \cref{sec:PTFIM_entanglement_clusters}. Nonetheless, taken as a whole we believe these results provide reasonable evidence to suggest that if the critical circuit dynamics has a simple geometric map to percolation, as in the PTFIM, then we should expect the scaling of the entanglement clusters to be controlled by surface exponents of $(d+1)$-dimensional percolation.

In fact, we will see that the critical properties of the entanglement clusters in the steady-state of the random Clifford circuits scale with exponents quite distinct from those of surface $(d+1)$-dimensional percolation. Several of them are controlled by exponents close to those of bulk $d$-dimensional percolation, but it is possible this could be a coincidence. We offer two possible interpretations of these results. First, this could be further evidence that the measurement-induced transition in random Clifford circuits on qubits is in a distinct universality class to percolation, which is the conclusion of several recent studies~\cite{zabaloCriticalPropertiesMeasurementinduced2020,liConformalInvarianceQuantum2020,zabaloOperatorScalingDimensions2021}. Second, lessons from Haar-random circuits \cite{jianMeasurementinducedCriticalityRandom2020} suggest that, even if a map to percolation does exist in certain limits, it may be highly non-trivial, and in particular may not have a simple geometric interpretation as for the PTFIM and the Hartley entropy in Haar circuits~\cite{skinnerMeasurementInducedPhaseTransitions2019}. As a consequence it is less obvious that the critical properties of the entanglement clusters in random Clifford circuits should be controlled by the surface exponents $\beta_{s}$ and $\gamma_{1,1}$ that are relevant for models that do have a simple geometric map to percolation.

Before we go into more detail, we make a brief comment about notation. As we just discussed, in the absence of a simple geometric map to percolation, it is not obvious that the mean and largest cluster sizes should be controlled by the surface exponents $\gamma_{1,1}$ and $\beta_{s}$ as they are in the PTFIM. For this reason we will label exponents for the entanglement clusters with the subscript $ec$, and do not claim that they should necessarily match the exponents $\gamma_{1,1}$ and $\beta_{s}$ in all models.

To find the entanglement clusters, we employ a breadth-first search on the graph storing the steady state~\cite{cormenIntroductionAlgorithms2009}. \cref{fig:mean_cluster_size} shows the behavior of the average cluster size $\overline{s} = \sum_{s} n_{s} s^{2} / \sum_{s'} n_{s'} s'$, where the cluster number $n_{s}$ is the number of clusters of size $s$ normalized by the system volume $L^{d}$. Note that this quantity measures the average cluster size if \textit{sites} are randomly selected with equal probability---if instead \textit{clusters} are randomly selected with equal probability then the corresponding average is $\sum_{s} n_{s} s / \sum_{s'} n_{s'}$. Assuming critical scaling of the form $\overline{s} \sim L^{\gamma_{ec}/\nu}$, the inset to \cref{fig:mean_cluster_size}a shows a log-log plot of this quantity for 1+1D Clifford circuits, with a fitted exponent of $\gamma_{ec}/\nu \approx 0.95(1)$ shown by the solid red line, close to the value of $\gamma/\nu = 1$ for 1D bulk percolation~\cite{staufferIntroductionPercolationTheory2018}, and far from the value $\gamma_{1,1}/\nu = 1/3$ for surface 2D percolation~\cite{binderCriticalBehaviourSurfaces1983}. The analogous plot for 2+1D Clifford circuits is shown in the inset to \cref{fig:mean_cluster_size}b, where the fitted exponent $\gamma_{ec}/\nu \approx 1.84(2)$ is close to the value $\gamma/\nu = 43/24 \approx 1.79$ for bulk 2D percolation, and far from the value $\gamma_{1,1}/\nu \approx 0.049$ for surface 3D percolation~\cite{binderCriticalBehaviourSurfaces1983,dengSurfaceCriticalPhenomena2005}.

\begin{figure}[t]
    \centering
    \includegraphics[width=\columnwidth]{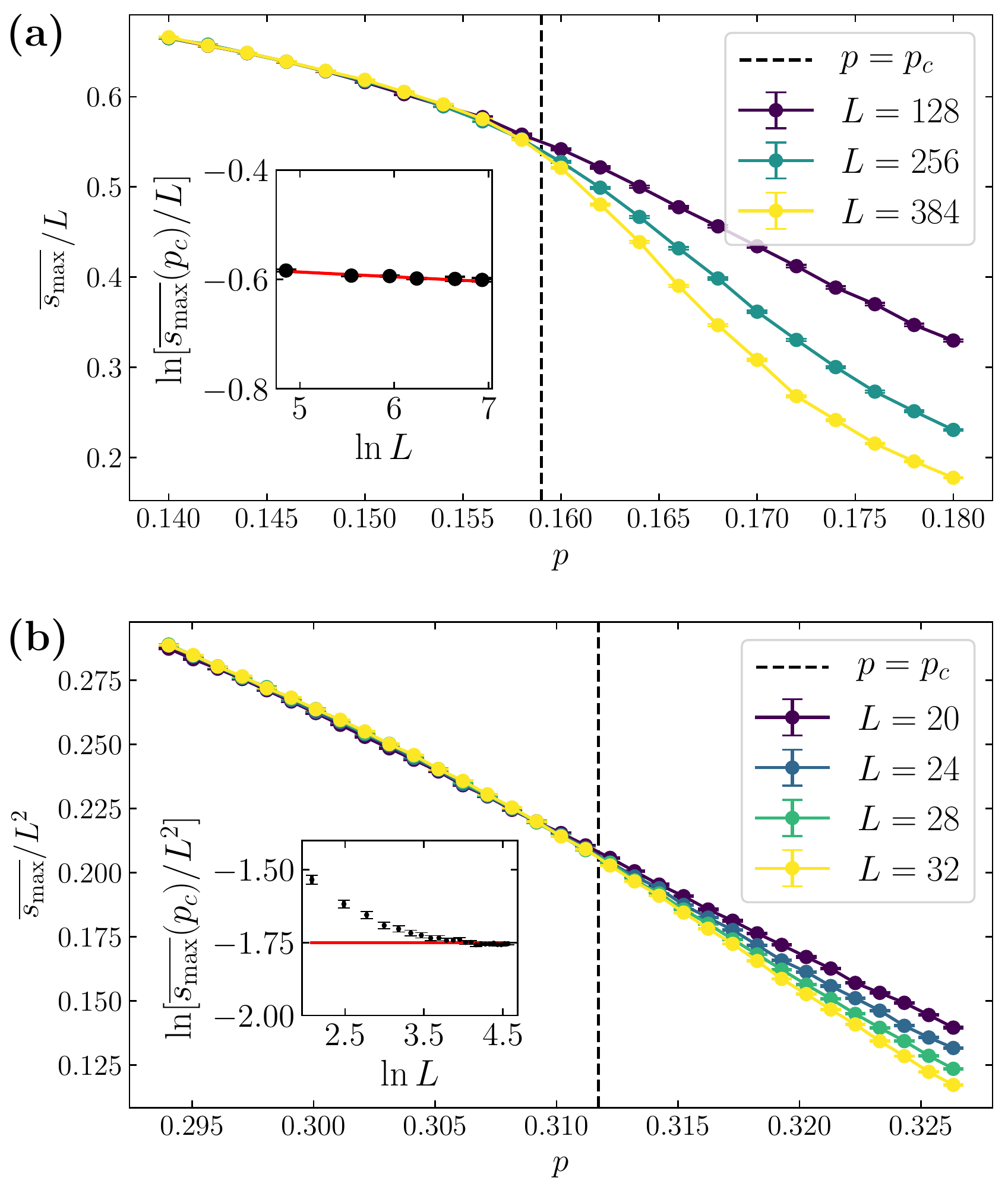}
    \caption{The average size $\overline{s_{\mathrm{max}}}$ of the largest entanglement cluster in the steady state for \textbf{(a)} 1+1D and \textbf{(b)} 2+1D Clifford circuits. The insets show log-log plots of this quantity at $p=p_{c}$, with the behavior well described by the power law $\overline{s_{\mathrm{max}}}(p_{c}) / L^{d} \sim L^{-\beta_{ec}/\nu}$, where $\beta_{ec}/\nu = -0.009(2)$ for 1+1D and $\beta_{ec}/\nu = 0.00(2)$ for 2+1D (power-law fits shown in solid red). Note there are strong finite-size effects in 2+1D, so there the fit is only to sizes $L \geq 40$.}
    \label{fig:largest_cluster_size}
\end{figure}

\begin{figure}[t]
    \centering
    \includegraphics[width=\columnwidth]{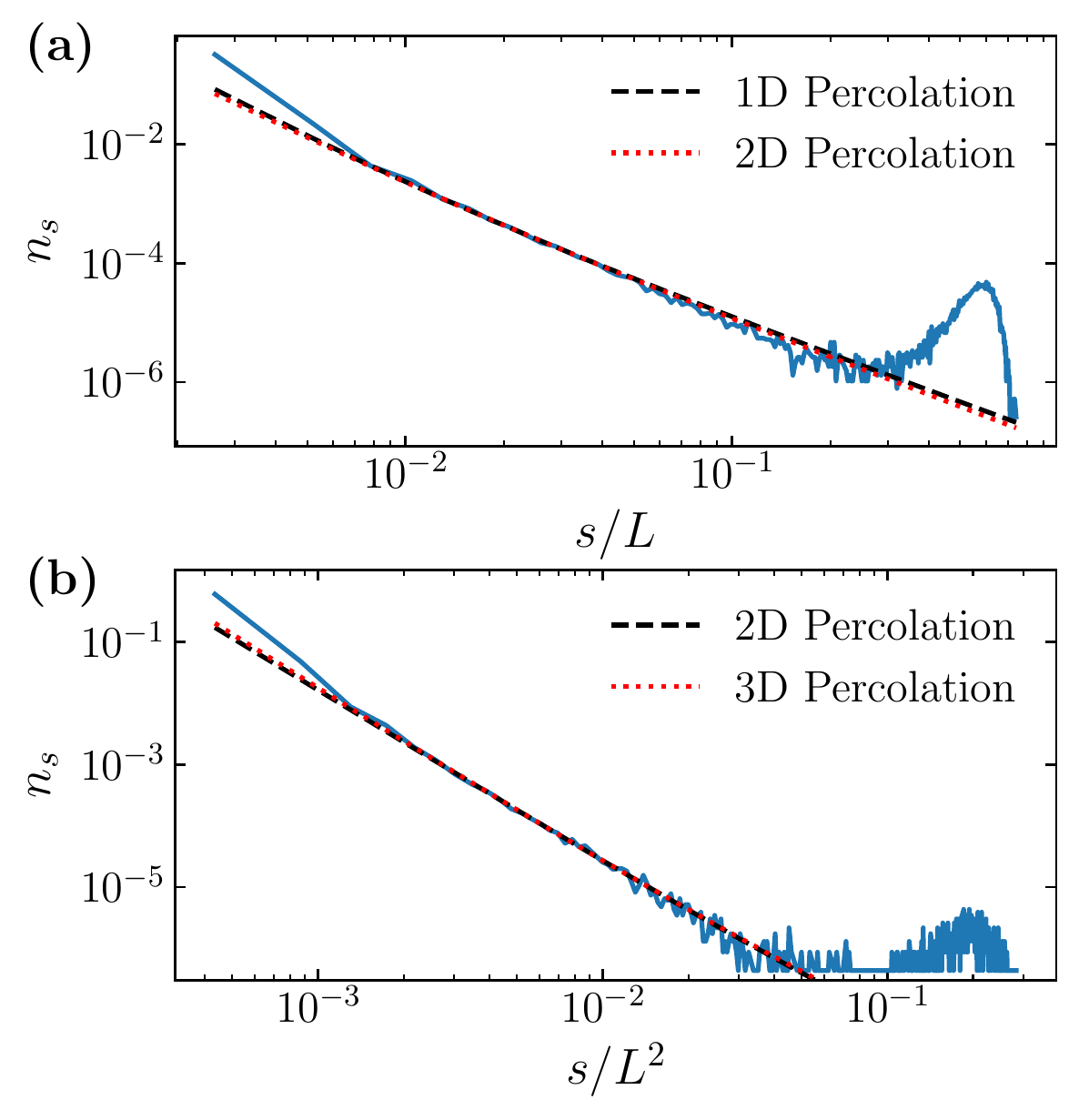}
    \caption{Distribution function $n_{s}$ of the entanglement cluster sizes $s$ in the $p=p_{c}$ steady state for \textbf{(a)} 1+1D and \textbf{(b)} 2+1D Clifford circuits, with system sizes $L=348$ and $L^{2} = 48^{2}$ respectively. For $1 \ll s \ll L^{d}$, the probability distribution follows a power-law distribution $n_{s} \sim s^{-\tau}(c_{0}+c_{1}s^{-\Omega})$ with the leading-order correction to scaling controlled by the exponent $\Omega$. The dashed and dotted lines show fits using the exponents from $d$- and $(d+1)$-dimensional percolation respectively. The peak at large $s$ corresponds to the percolating cluster of size $\mathcal{O}(L^{d})$ present for $p \leq p_{c}$.}
    \label{fig:cluster_size_distribution}
\end{figure}

\cref{fig:largest_cluster_size} shows the average over circuit realizations of the size $s_{\mathrm{max}}$ of the largest steady-state cluster in each realization, as a fraction of system size. This is a measure of the surface fractal dimension $d_{f}$ of the infinite cluster since by definition $s_{\mathrm{max}} \sim L^{d_{f}} \sim L^{d - \beta_{s}/\nu}$. The inset to \cref{fig:largest_cluster_size}a shows a log-log plot of $\overline{s_{\mathrm{max}}}(p_{c}) / L^{d} \sim L^{-\beta_{ec}/\nu}$ for 1+1D Clifford circuits, which is well described by the fitted exponent $\beta_{ec}/\nu \approx -0.009(2)$. This is close to the value $\beta/\nu = 0$ for 1D bulk percolation, and far from the exponent $\beta_{s}/\nu = 1/3$ for surface 2D percolation. In 2+1D, we find that there are significant finite-size effects affecting the scaling of the largest cluster size. For small system sizes, $L \lessapprox 32$, the power-law exponent is close to the bulk 2D percolation exponent $\beta/\nu = 5/48 \approx 0.10$, but this appears to be a finite-size effect. At larger system sizes the exponent saturates to approximately zero, with the fitted value $\beta_{ec}/\nu \approx 0.00(2)$, which is very far from the surface 3D percolation exponent of $\beta_{s}/\nu \approx 0.97$~\cite{dengSurfaceCriticalPhenomena2005}.

Finally, in \cref{fig:cluster_size_distribution} we show the distribution $n_{s}$ of all cluster sizes $s$, which at $p=p_{c}$ and for $1 \ll s \ll L^{d}$ follows a power-law $n_{s} \sim s^{-\tau}(c_{0} + c_{1} s^{-\Omega} + \cdots)$, with the leading-order correction to scaling controlled by the exponent $\Omega$. A comment on this scaling form is necessary if we are to make a comparison with 1D percolation. As noted above, for 1D percolation the critical probability is $p_{c} = 1$. This has the consequence that, strictly at $p=p_{c}$, there is only a single cluster which covers the whole system, $s_{\mathrm{max}} = L$, so for cluster sizes $s < s_{\mathrm{max}}$ the cluster number $n_{s} = 0$. Nonetheless, one can meaningfully define the Fisher exponent $\tau$ by analyzing the behavior of $n_{s}$ for $p < p_{c}$, where one finds $\tau = 2$ for 1D percolation. However, a key difference between the 1+1D hybrid quantum circuits we study and 1D percolation is that for the quantum circuits, $1-p_{c} \approx 0.84$ is different from unity, so there is still randomness at the critical point, and thus we can observe a full distribution of cluster sizes. This provides justification for continuing to use the scaling form $n_{s} \sim s^{-\tau}(c_{0} + c_{1} s^{-\Omega} + \cdots)$ to describe the cluster distribution function in 1+1D hybrid circuits.

In this case, it is harder to distinguish the behavior of $d$- and $(d+1)$-dimensional percolation, since the exponents for the leading term, $\tau_{\mathrm{1D}} = 2$, $\tau_{\mathrm{2D}} = 187/91 \approx 2.05$~\cite{staufferIntroductionPercolationTheory2018} and $\tau_{\mathrm{3D}} \approx 2.19$~\cite{xuSimultaneousAnalysisThreedimensional2014}, are all quite similar in magnitude. Indeed both $\tau_{\mathrm{1D}}$ and $\tau_{\mathrm{2D}}$ provide a reasonable fit to our 1+1D data (see \cref{fig:cluster_size_distribution}a), and both $\tau_{\mathrm{2D}}$ and $\tau_{\mathrm{3D}}$ provide a reasonable fit to our 2+1D data (see \cref{fig:cluster_size_distribution}b). An independent statistical bootstrap analysis~\cite{efronIntroductionBootstrap1994} gives the exponents $\tau \approx 2.04$ and $\Omega \approx 0.15$ in 1+1D, and $\tau \approx 1.98$ and $\Omega \approx 1.04$ in 2+1D. However, it is hard to call these values physically meaningful, since allowing for variation in the scaling correction exponent $\Omega$ provides considerable freedom to optimize the quality of the fit. What is at least clear is that the Fisher exponent $\tau$ is close to values predicted by percolation theory in \textit{low dimensions}, since our fitted values are far from the mean-field value $\tau = 2.5$.

We conclude this section by noting that the entanglement cluster distribution is qualitatively similar to the stabilizer length distribution (SLD) introduced by Li, Chen and Fisher in Ref.~\cite{liMeasurementdrivenEntanglementTransition2019}. Indeed, both have a power-law tail, and a volume-law peak which disappears upon entering the area-law phase. Furthermore, at criticality the exponent $\tau$ of the power-law tail is close to 2 in both cases. In 1+1D the SLD has the nice property that it can be used to calculate the entanglement entropy itself---for example, a power-law exponent of 2 gives rise to a subleading $\log{L}$ contribution to the entanglement entropy. However, it is not clear how to generalize the SLD to higher dimensions in a way that preserves this ability to calculate the entanglement entropy from the analogous `stabilizer volume distribution'. From the entanglement cluster distribution we analyze here, it is possible to calculate the entanglement entropy provided one makes certain simplifying assumptions about the fractal structure of the entanglement clusters, but we defer further analysis of this link to future work.

\section{Discussion}

We have provided an extensive study of the critical properties of the measurement-induced transition in 2+1D Clifford circuits. Analogously to the situation in 1+1D, we have found several bulk critical exponents which are within error-bars of those from 3D percolation, but there appear to be some differences in surface behavior. We should note that these critical exponent estimates should be treated with some amount of caution, especially for small system sizes, as conformal field theories with zero central charge (like those appearing in current theories of the 1+1D transition~\cite{jianMeasurementinducedCriticalityRandom2020}) can have logarithmic corrections to scaling~\cite{gurarieConformalAlgebrasTwodimensional2002, cardyLogarithmicConformalField2013}, which could result in systematic errors.

Nonetheless, focusing on this surface behavior, we studied the critical scaling of entanglement clusters in the steady state, and found that --- in contrast to models with a simple geometric map to percolation --- Clifford circuits have entanglement cluster exponents which differ significantly from those of surface percolation. We take this as evidence that in 1+1D and 2+1D the measurement-induced transition in qubit Clifford circuits is in a distinct universality class from percolation.

Presumably the entanglement clusters are governed by surface exponents of the as yet unknown $(d+1)$-dimensional statistical mechanical model applicable to Clifford circuits. It remains a significant question why the bulk exponents of this model look so much like those of percolation, even though this system is far from where the percolation picture should be applicable. There have been recent developments in the machinery required to average over random Clifford unitaries~\cite{grossSchurWeylDuality2021}, which should prove helpful in developing this statistical mechanical model. However, the reduced structure relative to Haar-random unitaries makes it less obvious how to perform the replica limit required to give the correct critical physics.

We have also shown the coincidence of the purification transition and the entanglement transition in 2+1D. This may at first be surprising, given that the entanglement transition concerns spatial correlations between equal-time wavefunctions, while the purification transition concerns correlations in time of a non-local quantity. The results in this paper indicate that these two transitions may coincide in all dimensions. One possible explanation for this could be the conjecture of Ref.~\cite{liConformalInvarianceQuantum2020} that the non-unitary nature of the dynamics results in the real time coordinate in $d$ spatial dimensions acting as imaginary time in the corresponding $(d+1)$-dimensional statistical mechanical model. In this sense space and time may become symmetric, so the coincidence of the entanglement transition and the purification transition would be less surprising. The coincidence of these transitions and our entanglement cluster analysis also suggest a way to investigate connections with quantum error-correction---the emergence of the critical entanglement cluster can be seen as the germination of the quantum error-correcting code that characterizes the stability of the volume-law phase.

Moving into higher dimensions raises several questions. One interesting direction is that of `measurement-protected order'~\cite{sangMeasurementProtectedQuantum2020,lavasaniMeasurementinducedTopologicalEntanglement2020}, analogous to the `localization-protected order' afforded by many-body localization (MBL) \cite{huseLocalizationprotectedQuantumOrder2013,bauerAreaLawsManybody2013}. It is tempting to view the area-law side of the measurement-induced transition as a `trivial' phase, but recent work has demonstrated that there can be stable symmetry-protected topological (SPT) order in the area-law phase, motivated by comparisons with the area-law ground states of gapped Hamiltonians. However, it is only in dimensions $d \geq 2$ that true topological order can exist~\cite{schuchClassifyingQuantumPhases2011}, so it would be interesting to see if non-trivial topological order could be realized in the steady states of 2+1D hybrid quantum circuits. There is also the question of which types of order can be stabilized by measurements. There are significant constraints on possible phases stabilized by MBL: non-Abelian symmetries are forbidden~\cite{potterSymmetryConstraintsManybody2016}, for example, as well as chiral order~\cite{potterProtectionTopologicalOrder2015}. It is also possible that true MBL does not exist in $d > 1$~\cite{deroeckStabilityInstabilityDelocalization2017}. It is an important topic for future research to determine which restrictions, if any, are applicable to measurement-protected order. This may allow for considerably more freedom in the more general paradigm of understanding and classifying non-equilibrium phases of matter.

\begin{acknowledgments}
We thank Brian Skinner, Matthew Fisher, Michael Gullans, David Huse, Keith De'Bell and Robert Ziff for helpful discussions. A.P.\ and M.S.\ were funded by the European Research Council (ERC) under the European Union's Horizon 2020 research and innovation programme (grant agreement No.\ 853368). O.L.\ was supported by the Engineering and Physical Sciences Research Council [grant number EP/L015242/1].
\end{acknowledgments}

\bibliography{bibliography}

\clearpage

\onecolumngrid
\appendix

\renewcommand\thefigure{\thesection\arabic{figure}}    
\setcounter{figure}{0}  

\section{Alternative scaling forms for $I_{3}$}
\label{sec:alternative_scaling}

\begin{figure}[!t]
    \centering
    \includegraphics[width=\columnwidth]{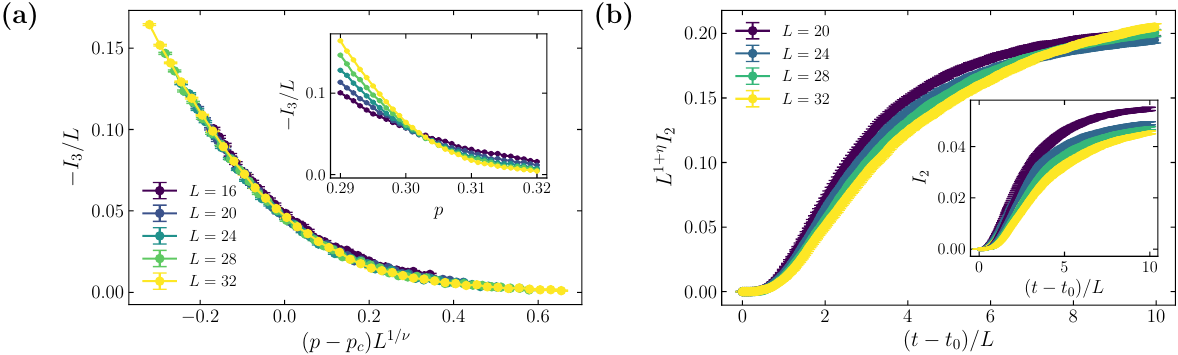}
    \caption{\textbf{(a)} The steady-state values of $I_{3}/L$ as a function of $(p-p_{c})L^{1/\nu}$, where $p_{c} \approx 0.303$ and $\nu \approx 1.07$. The inset shows the uncollapsed data. This dataset consists of 50,000 circuit realizations. \textbf{(b)} Analogous to \cref{fig:eta_exponent}a, except performed at the alternative critical point $p_{c} \approx 0.303$ estimated from the data collapse of $I_{3}/L$. The main plot shows the `optimal' collapse at $\eta = -0.57$ as determined by minimizing the cost function in \cref{sec:cost_function}, but this clearly does not produce a good data collapse.}
    \label{fig:A1}
\end{figure}

In this section we detail some evidence against the hypothesis that $I_{3} \sim \mathcal{O}(L)$ at $p=p_{c}$. Finite-size scaling of $I_{3} / L$ results in the critical point $p_{c} \approx 0.303$ with $\nu \approx 1.07$ (see \cref{fig:A1}a). However, if we attempt to use this critical point to estimate other critical exponents from standard finite-size scaling arguments, we are unable to obtain a good data collapse, indicating the absence of scaling behavior. For example, to extract the anomalous scaling exponent $\eta$,  we follow the procedure detailed in \cref{sec:purification}, where $\eta$ is chosen to optimize the data collapse of the dynamics of the mutual information between two ancilla qubits. Whereas this was possible for the critical point $p_{c} \approx 0.312$ obtained from finite-size scaling of $I_{3}$ (see \cref{fig:eta_exponent}a), for the purported critical point $p_{c} \approx 0.303$ from $I_{3} / L$ scaling, there was not a value of $\eta$ for which a good data collapse was possible (see \cref{fig:A1}b). Moreover, the data collapse in \cref{fig:A1}a is of visibly worse quality than the excellent collapse in \cref{fig:I3_transition}. We also see in \cref{sec:purification} that the purification transition seems to coincide with the critical point $p_{c} \approx 0.312$ from $I_{3}$ scaling, with a dynamical critical exponent $z \approx 1$ indicating the emergence of conformal symmetry. Given that these facts mirror the situation in 1+1D, this provides further \textit{a posteriori} justification for the scaling $I_{3} \sim \mathcal{O}(1)$ at criticality.

\section{Details of the finite-size scaling}
\label{sec:cost_function}

\begin{figure}[!t]
    \centering
    \includegraphics[width=\columnwidth]{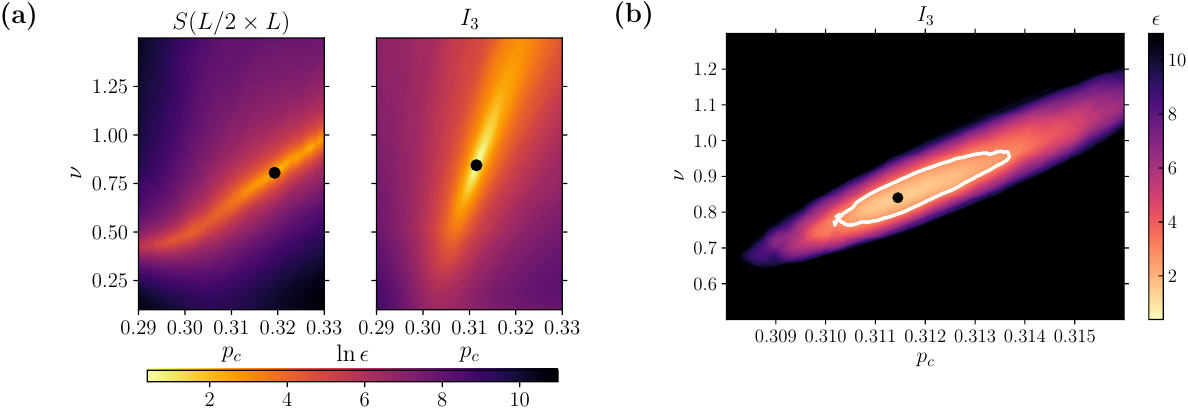}
    \caption{\textbf{(a)}~The logarithm of the cost function $\epsilon$ measuring the quality of the data collapse for different values of $p_{c}$ and $\nu$, compared between two possible indicators of the entanglement transition: the half-plane entanglement entropy $S(L/2)$, and the tripartite information $I_{3}$. The black dots show the minimum of the cost function for each indicator. See the appendix for a definition of the cost function $\epsilon$. \textbf{(b)}~A linear-scale close-up of the cost function for the $I_{3}$ data collapse around the estimated critical point, which is indicated by the black dot. The white line indicates the boundary of the region for which the cost function is less than 2 times its minimum value; this is the region from which the error bars are calculated. At the estimated critical point the cost function attains the value $\epsilon = 1.47$, close to the optimal value $\epsilon \approx 1$. }
    \label{fig:B2}
\end{figure}

To perform the data collapses, we use a cost function $\epsilon(p_{c}, \nu)$ which uses linear interpolation to find the parameters $(p_{c}, \nu)$ which cause the data to best collapse on to a single curve~\cite{kawashimaCriticalBehaviorThreeDimensional1993,zabaloCriticalPropertiesMeasurementinduced2020}. In more detail, given a set of parameters $(p_{c}, \nu)$, for each value of $p$ and $L$ we create an $x$-value $x \coloneqq (p-p_{c})L^{1/\nu}$, with a corresponding $y$-value $y(p,L)$ and error $d(p,L)$. We then sort the triples $(x_{i},y_{i},d_{i})$ according to their $x$-values, and evaluate the cost function

\begin{equation}
    \epsilon(p_{c},\nu) \coloneqq \dfrac{1}{n-2} \sum_{i=2}^{n-1} w(x_{i},y_{i},d_{i} | x_{i-1},y_{i-1},d_{i-1},x_{i+1},y_{i+1},d_{i+1}),
\end{equation}
where $w(x_{i},y_{i},d_{i} | x_{i-1},y_{i-1},d_{i-1},x_{i+1},y_{i+1},d_{i+1})$ is defined as
\begin{align}
   w &\coloneqq \left( \dfrac{y - \bar{y}}{\Delta(y-\bar{y})} \right)^{2},\\
   \bar{y} &\coloneqq \dfrac{(x_{i+1} - x_{i})y_{i-1} - (x_{i-1} - x_{i})y_{i+1}}{x_{i+1} - x_{i-1}},\\
   \left| \Delta(y-\bar{y}) \right|^{2} &\coloneqq d_{i}^{2} + \left( \dfrac{x_{i+1} - x_{i}}{x_{i+1}-x_{i-1}} \right)^{2} d_{i-1}^{2} + \left( \dfrac{x_{i-1} - x_{i}}{x_{i+1} - x_{i-1}} \right)^{2} d_{i+1}^{2}.
\end{align}

The function $w$ measures the deviation of a point from the line obtained by a linear interpolation of its nearest neighbours, weighted by the errors in each data point. Values of $(p_{c},\nu)$ for which $\epsilon(p_{c},\nu) \approx 1$ are considered optimal.

As discussed in \cref{sec:entanglement_transition}, our finite-size scaling analysis yields the correlation length exponent $\nu \approx 0.85(9)$, which is significantly different to that observed in Ref.~\cite{turkeshiMeasurementinducedCriticalityDimensional2020}. We attribute this to the fact that we extract $\nu$ by a data collapse not of the half-plane entanglement, as in Ref.~\cite{turkeshiMeasurementinducedCriticalityDimensional2020}, but of the tripartite information, which coincides for different system sizes at the critical point and so provides a much more accurate estimator of the critical point. To further this point, we show in \cref{fig:B2}a a comparison of the cost function $\epsilon(p_{c},\nu)$ landscape in log scale between the half-plane entropy $S(L/2 \times L)$ and the tripartite information $I_{3}$. The entropy cost function plot shows a clear `ridge' region where $\epsilon$ is roughly constant, spanning the whole range of values of $p_{c}$ and with a large variation of $\nu$ along the ridge (see also Fig.\ 2 in the Erratum of Ref.~\cite{turkeshiMeasurementinducedCriticalityDimensional2020}). On the other hand, the $I_{3}$ cost function plot is much more localized around the estimated critical parameters, reaching a smaller value of $\epsilon$ than the entropy plot. This localization is less obvious viewed in log scale, but the log was necessary for a meaningful visual comparison of the cost function plots for the two indicators. \cref{fig:B2}b shows a linear-scale version of the cost function plot for $I_{3}$, which allows for a clearer visualization of the localization of the cost function minimum. The estimated critical point is indicated by the large black dot, while the surrounding white line gives the boundary of the region where the cost function is less than 2 times its minimum value, from which we calculate the error bars in $p_{c}$ and $\nu$. Notice that at the estimated critical point, the cost function reaches a value $\epsilon = 1.47$ close to 1, indicating a good-quality data collapse. 

Furthermore, a comment on the used system sizes is necessary. One could argue that the 4 subsystems used to calculate $I_3$ have the vertical dimension $L_y \leq 8$, which may be small enough to exhibit substantial finite-size effects, hindering our ability to properly locate the critical point. However, $I_3$ in 1+1D circuits shows almost no finite-size drift at criticality already for systems of size $L\geq 16$~\cite{zabaloCriticalPropertiesMeasurementinduced2020} (subsystems of size $\geq 4$). Using our data from \cref{fig:I3_transition}, one can assess that the crossings of $I_3$ exhibit no statistically significant drift above roughly $L \geq 16$, strongly implying little to no finite-size effects in $I_3$ at criticality for the system sizes considered. We also note that the data collapse is of exceptional quality, again strongly ruling out any substantial finite-size drifts.

\section{Critical entanglement dynamics}
\label{sec:appendix_entanglement_dynamics}

\begin{figure}[t]
    \centering
    \includegraphics[width=0.9\columnwidth]{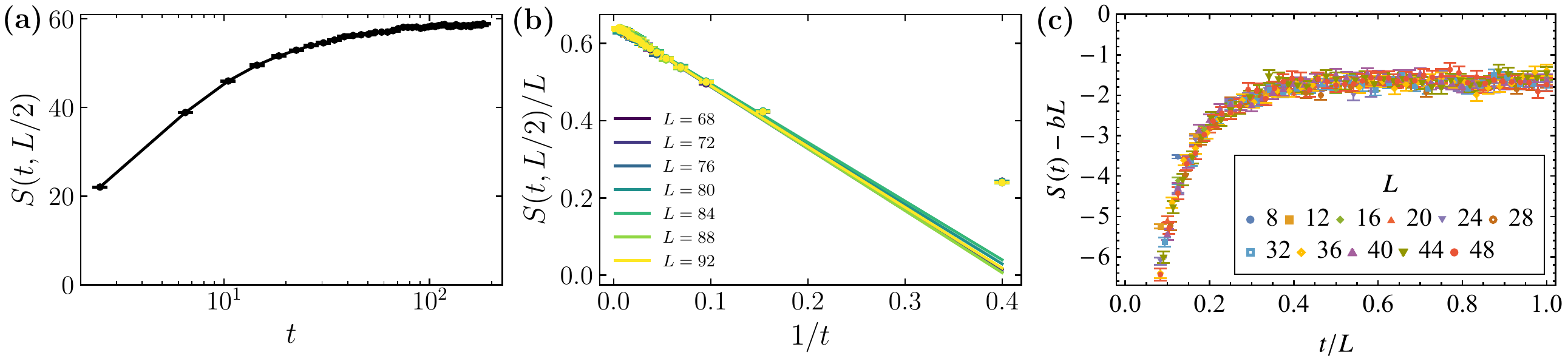}
    \caption{The dynamics of the half-plane von Neumann entropy at the critical point $p_{c}=0.312$ of the 2+1D Clifford model. \textbf{(a)} The data are not linear on a log scale, indicating that the entanglement growth is \textit{not} logarithmic in time (system size is $L=92$). \textbf{(b)} A plot of $S(t,L/2)/L$ as a function of $1/t$, where the linear trend provides support for the scaling $S(t,L) \sim L(1 - a/t)$. \textbf{(c)} Scaling collapse of $S(t) - b L$ vs $t/L$, with $b=0.685$ producing the best fit.} 
    \label{fig:critical_entanglement_dynamics}
\end{figure}

In \cref{fig:critical_entanglement_dynamics} we plot the dynamics of the half-plane von Neumann entropy at the 2+1D critical point $p_{c} = 0.312$. Because the entanglement is relatively small at the critical point, we are able to simulate a large system with linear size $L = 92$. In \cref{fig:critical_entanglement_dynamics}a the time axis is on a logarithmic scale, and we can see that the data do not appear linear on this scale, thereby demonstrating that the entanglement growth is not logarithmic in time. Note that we are plotting here the window-averaged entropy, averaged over a window of 4 timesteps, which is why there is not data at every timestep. This is to remove a periodicity effect related to how often the Clifford gates cross the cut used to define the entanglement entropy, as discussed in \cref{sec:method}A.

As we discuss in \cref{sec:entanglement_transition}, we instead argue that the entanglement growth scales as $S(t,L) = b L(1-a/t)$ in 2+1D, where $a,b$ are some $\mathcal{O}(1)$ constants. Evidence for this is shown in \cref{fig:critical_entanglement_dynamics}b, where the data appears approximately linear when plotted as a function of $1/t$. Note that the data appears linear on this scale, with the straight lines showing linear fits. The gradients and y-intercepts of these fits are approximately the same for different system sizes, supporting the idea that $a$ and $b$ are $\mathcal{O}(1)$ constants. Note that we only expect this scaling to hold for intermediate times, so there are some deviations from this behavior at early times. Finally, in \cref{fig:critical_entanglement_dynamics}c, we show a data collapse of $S(t) - b L$ vs $t/L$ with $b = 0.685$, supporting the scaling ansatz $S(t) - b L \sim f(t/L)$ consistent with a dynamical critical exponent of $z=1$.

\section{Entanglement clusters in the projective transverse field Ising model}
\label{sec:PTFIM_entanglement_clusters}

The projective transverse field Ising model (PTFIM) is a measurement-only model exhibiting an entanglement transition which is known to be in the percolation universality class \cite{langEntanglementTransitionProjective2020}. Conveniently, it also only involves Clifford operations, so can be simulated using the graph-state framework, and therefore provides a useful testbed for the scaling properties of the entanglement clusters we analyze in \cref{sec:entanglement_clusters}.

Referring the reader to Ref.~\cite{langEntanglementTransitionProjective2020} for the full details, the PTFIM is defined as follows. We define the model on a hypercubic lattice for simplicity. Each site of the lattice contains a spin. The model involves two types of measurements: on-site measurements of $\sigma^{x}$, and measurements of $\sigma^{z}\sigma^{z}$ for spins connected by an edge. The system is initialized in the product state $|+\rangle^{\otimes N}$, where $|+\rangle = (|0\rangle + |1\rangle)/\sqrt{2}$. Then, at each timestep, for each site $i$ assign the variable $x_{i} = 1$ with probability $p$ and $x_{i} = 0$ otherwise, and for each edge $e$ connecting spins $i$ and $j$, assign the variable $z_{e} = 1$ with probability $1-p$ and $z_{e} = 0$ otherwise. These variables determine the sites and edges on which the observables $\sigma^{x}_{i}$ and $\sigma^{z}_{i} \sigma^{z}_{j}$ are measured. The edge observables are measured first, followed by the site observables. On a $d$-dimensional hypercubic lattice, this process maps on to bond percolation on a $(d+1)$-dimensional hypercubic lattice.  

\begin{figure}[t]
    \centering
    \includegraphics[width=0.9\columnwidth]{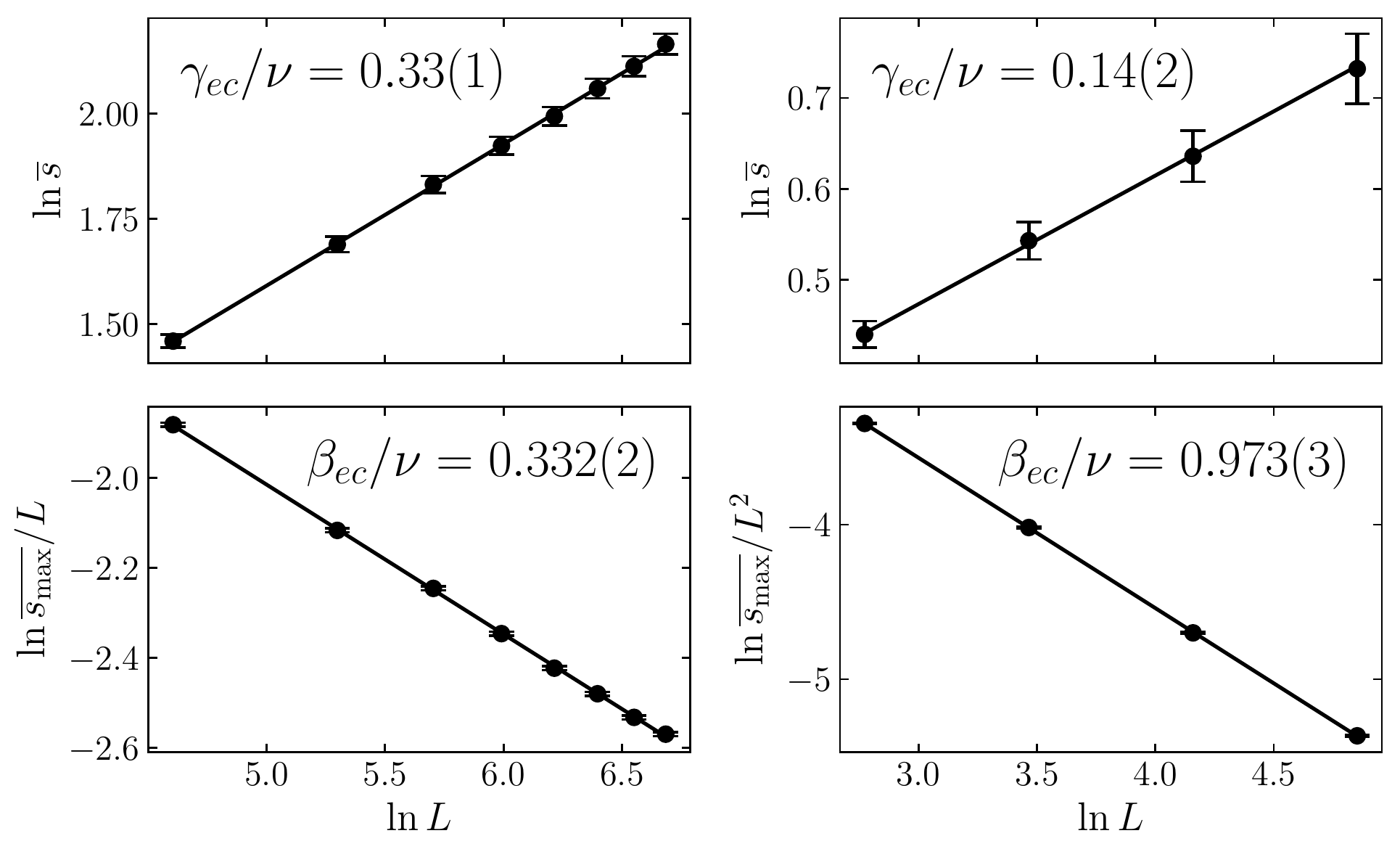}
    \caption{The mean cluster size $\overline{s}$ and largest cluster size $s_{\mathrm{max}}/L^{d}$ for the projective transverse field Ising model in 1+1D (left column) and 2+1D (right column). These should scale as $\overline{s} \sim L^{\gamma_{1,1}/\nu}$ and $s_{\mathrm{max}} / L^{d} \sim L^{-\beta_{s}/\nu}$ respectively. The critical exponents are all close to the corresponding surface critical exponents of percolation in $(d+1)$-dimensions, with the exception of the mean cluster size in 2+1D, as we discuss in the main text.}
    \label{fig:projective_TFIM_combined_plots}
\end{figure}

\begin{figure}[t]
    \centering
    \includegraphics[width=0.9\columnwidth]{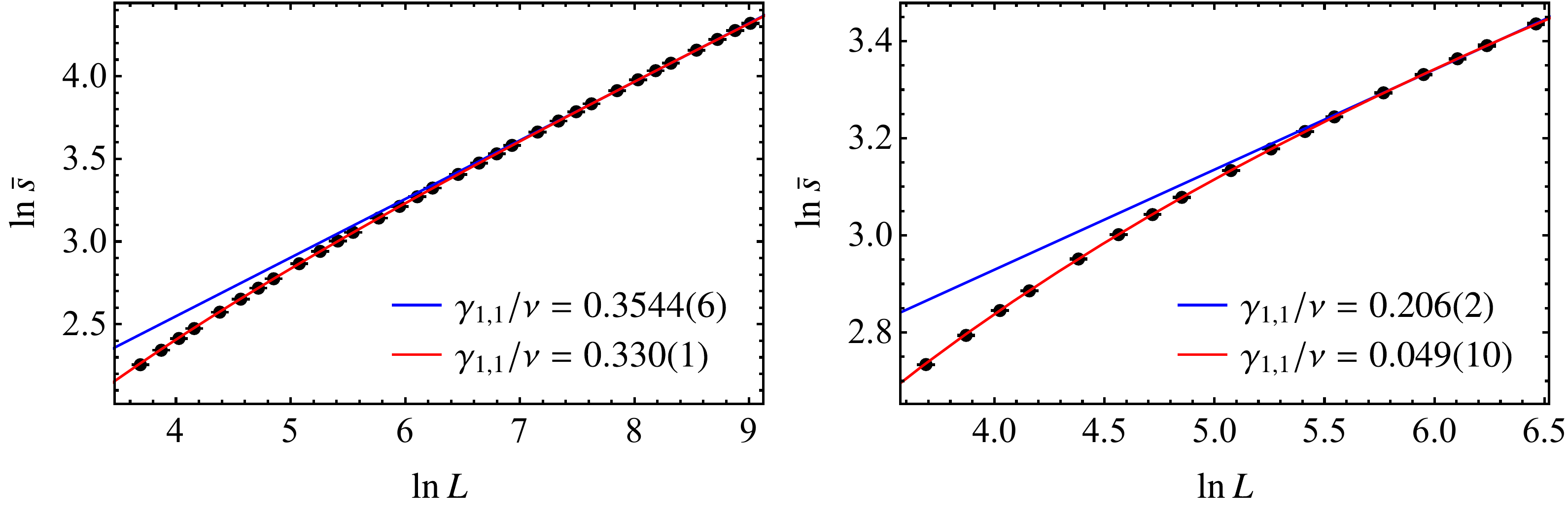}
    \caption{The mean surface cluster size $\overline{s}$ for the site percolation on a 2D square lattice (left) and a 3D simple cubic lattice (right). Blue line is a fit to $\overline{s} = a L^{\gamma_{1,1}/\nu}$ for the largest system sizes, while the red line includes a constant correction to scaling, $\overline{s} = a L^{\gamma_{1,1}/\nu} + b$. Corresponding estimates of $\gamma_{1,1}/\nu$ are given in the legend.}
    \label{fig:percolation_exponents}
\end{figure}

As previously, we focus on two properties of the surface clusters: the largest cluster size $s_{\mathrm{max}}$, and the mean cluster size $\overline{s}$. In a system with $d$ spatial dimensions and linear size $L$, these should scale as $s_{\mathrm{max}} / L^{d} \sim L^{-\beta_{s}/\nu}$ and $\overline{s} \sim L^{\gamma_{1,1}/\nu}$ respectively. Our results for the PTFIM in 1+1D and 2+1D are shown in the left and right columns of \cref{fig:projective_TFIM_combined_plots}, where we perform simulations up to $L=800$ and $L=128$ respectively. In 1+1D, the resulting exponents for the entanglement clusters are $\gamma_{ec}/\nu = 0.33(1)$ for the mean cluster size and $\beta_{ec}/\nu = 0.332(2)$ for the largest cluster size. These are very close to the corresponding surface exponents for 2D percolation, $\gamma_{1,1}/\nu = 1/3$ and $\beta_{s}/\nu = 1/3$. In 2+1D, the extracted exponent for the largest cluster size is $\beta_{ec}/\nu = 0.973(3)$, which is very close to the exponent $\beta_{s}/\nu \approx 0.9754(4)$ for 3D percolation~\cite{dengSurfaceCriticalPhenomena2005}. For the mean cluster size in 2+1D, the situation is less clear. We extract an exponent for the entanglement clusters of $\gamma_{ec} / \nu = 0.14(2)$. The exponent $\gamma_{1,1}$ does not appear to be well documented for 3D percolation, however, from the scaling relation $\gamma_{1,1}/\nu = d - 1 - 2\beta_{s}/\nu$ \cite{DeBell1980, binderCriticalBehaviourSurfaces1983} we estimate the value $\gamma_{1,1}/\nu = 0.0492(8)$, which is not compatible within error bars of the exponent $\gamma_{ec}/\nu$. Nonetheless, it is very likely that there are large finite size effects for this exponent --- we have performed percolation simulations (see \cref{fig:percolation_exponents}) to reproduce the quoted value for $\gamma_{1,1}/\nu$, and found that we had to be very careful with the subleading corrections to scaling in order to get the correct exponent, even up to surprisingly large system sizes ($L \le 640$). Without accounting for the corrections, we obtain a larger exponent, $\gamma_{1,1}/\nu \sim 0.206(2)$, while including a constant correction gives $\gamma_{1,1}/\nu \sim 0.049(10)$, a value close to the expectation from the scaling relation. For the 2+1D Clifford circuit we have simulated up to $L=128$ at criticality, but it is possible that there are still significant finite size corrections to $\gamma_{ec}/\nu$ that are not captured by the statistical error bars we quote here. It, however, needs to be noted that there are relatively large error bars on $\overline s$ for PTFIM in 2+1D which could conceal finite size effects, while the corresponding results for Clifford circuit have smaller error bars and seem to exhibit small finite size effects.

\end{document}